\documentclass[twocolumn]{IEEEtran}

\usepackage{amsmath}
\usepackage{amsfonts}
\usepackage{mathrsfs}
\usepackage{pifont}
\usepackage{amssymb}
\usepackage{verbatim}
\usepackage{upgreek}
\usepackage[final]{graphicx}
\usepackage{booktabs}
\usepackage{bm}
\usepackage{setspace}
\usepackage{url}
\usepackage[utf8]{inputenc}
\usepackage{algorithmic}
\usepackage{algorithm}
\usepackage[overload]{empheq}
\usepackage{fancyhdr}
\usepackage{cite}

\usepackage{xcolor}

\usepackage{amsthm}

\ifCLASSOPTIONcompsoc
    \usepackage[caption=false, font=normalsize, labelfont=sf, textfont=sf, subrefformat=parens,labelformat=parens]{subfig}
\else
\usepackage[caption=false, font=footnotesize, subrefformat=parens,labelformat=parens]{subfig}
\fi

\usepackage{etoolbox}
\robustify\subref

\usepackage{graphicx}
\usepackage{grffile}
\usepackage[shortlabels]{enumitem}

\newcommand{\bSigma}{\boldsymbol{\Sigma}}

\newcommand{\bGamma}{\boldsymbol{\Gamma}}

\newcommand{\ba}{\boldsymbol{a}}

\newcommand{\bp}{\boldsymbol{p}}

\newcommand{\bs}{\boldsymbol{s}}

\newcommand{\bv}{\boldsymbol{v}}

\newcommand{\bx}{\boldsymbol{x}}

\newcommand{\bA}{\boldsymbol{A}}

\newcommand{\bD}{\boldsymbol{D}}
\newcommand{\bE}{\boldsymbol{E}}

\newcommand{\bS}{\boldsymbol{S}}

\newlength\myindent
\usepackage{cite}
\usepackage{amsmath,amssymb,amsfonts}
\usepackage{algorithmic}
\usepackage{graphicx}
\usepackage{textcomp}
\usepackage{xcolor}
\def\BibTeX{{\rm B\kern-.05em{\sc i\kern-.025em b}\kern-.08em
		T\kern-.1667em\lower.7ex\hbox{E}\kern-.125emX}}

\title{Reconfigurable Intelligent Surfaces for N-LOS Radar Surveillance}

\author{Augusto Aubry, \emph{Senior Member, IEEE}, Antonio De Maio, \emph{Fellow, IEEE}, and Massimo Rosamilia, \emph{Student Member, IEEE}
	\thanks{Augusto Aubry, Antonio De Maio (Corresponding Author), and Massimo Rosamilia are with Universit\`a degli Studi di Napoli ``Federico II'', DIETI, Via Claudio 21, I-80125 Napoli, Italy. E-mail: augusto.aubry@unina.it, ademaio@unina.it, massimo.rosamilia@unina.it.}
}

\begin{document}
	\maketitle

	\begin{abstract}
		This paper deals with the use of Reconfigurable Intelligent Surfaces (RISs) for radar surveillance in Non-Line Of Sight (N-LOS) scenarios. First of all, the geometry of the scene and the new system concept is described with emphasis on the required operative modes and the role played by the RIS. Then, the specific radar equation (including the RIS effect) is developed to manage the coverage requirements in the challenging region where the LOS is not present. Both noise and clutter interference cases (pulse length-limited and beamwidth-limited surface clutter as well as volume clutter) are considered.
		Hence, a {digression on} the use of the radar timeline for the new operative mode is presented together with {the data acquisition procedure and the} resolution issues for the range, azimuth, and Doppler domains.
		Finally, the interplay {among} the system parameters and, in particular, those involving the RIS is discussed and analyzed via numerical simulations.
	\end{abstract}
	
	
	% \IEEEpeerreviewmaketitle

	\section{Introduction}	
	Reconfigurable Intelligent Surfaces (RISs) are a novel and promising technology which is receiving growing interest in recent years especially for next generation communication ({e.g.,} beyond 5G and future 6G) and {sensing} systems~\cite{8466374,  8796365, 9124704, 9206044, 9226135, 9328302, 9359529, 9361184,   9364358}.
	RISs are man-made surfaces capable of varying the electric field distribution of the impinging signals, i.e., phase, amplitude, frequency, and polarization features, via appropriate electronic controls. Therefore, they pave the way to the ``lifelong dream'' of eliminating the Radio Frequency (RF) propagation medium randomness and avoiding the resulting deleterious effects, by means of a wisely designed electromagnetic waves interaction. Otherwise stated, they lay the ground to the paradigm of {\em smart radio environments}~\cite{9140329}.
	
	The use of RIS was firstly proposed {for communication purposes} in~\cite{6206517}, but it {has} gained serious attention from the research community since its employment as a phase-shifter in~\cite{7510962}. 
	{Nowadays}, excellent tutorial papers are available in open literature about diverse aspects connected with RIS technology and usage. For instance, hardware issues are addressed in~\cite{9140329} and~\cite{10.1002/adom.202000783}; applications for communication purposes are addressed in~\cite{9326394, 9086766, 8796365}; localization aspects are considered in~\cite{9215972} whereas a signal processing perspective is provided in~\cite{bjornson2021reconfigurable}. They also contain a selection of references which {extensively} span the state of the art for the current literature.
	
	As to the radar research field, applications of the RIS concept are mainly related 
	to the use of meta-surfaces for target Radar Cross Section (RCS) reduction~\cite{201600202, Gao2015}.
	Recently, other specific radar signal problems have also been tackled via RIS. Precisely, in~\cite{9364358} the RIS is integrated in a Dual-function Radar and Communication (DRC) system to improve radar detection performance in a crowded area with severe path loss. In~\cite{9328302} and~\cite{9361184}, the RIS is {employed} for colocated Multiple-Input Multiple-Output MIMO and Distributed MIMO (DMIMO) radar systems, respectively. The last reference capitalizes the presence of the {intelligent surface} which, via a specific coordination, aims at increasing the signal power from the prospective target.

	{At large the new technology appears promising to aid a multitude of radar applications. Leveraging this last observation, the present paper is aimed at exploring the usage of RIS for} radar detection in the absence of a direct path between radar and target. {This} is of primary interest in many practical civilian and military scenarios. {In fact}, the so called ``around-the-corner radar'' (see~\cite{5682038,iet-rsn.2014.0337, 6212140, 8640240} and references therein) {can be regarded as} a present-day and very challenging issue for security concerns especially in urban environments which are densely populated by small moving targets such as pedestrians and small Unmanned Aerial Vehicles (UAV). It is quite common that such kind of {targets} might fall in a shadow area (caused for instance by a tall building) blocking the LOS. As a consequence, the signal and data processing require adaptation to this unconventional propagation situation and new techniques {are} demanded to handle the problem. The {major strategies} developed in the open literature {rely} on natural multipath {capitalization} and/or on the use of radar networks. For instance, in~\cite{8640240} detection and localization of targets in Non-Line Of Sight (N-LOS) areas with a single portable radar via multipath exploitation is addressed. Two algorithms (which process the multipath returns) are proposed to detect the target and estimate its N-LOS position. In~\cite{10.1117/12.2325637}, a system incorporating distinct radar sensors (with communication capabilities) is developed for UAV detection in a urban environment. It includes small radar sensor nodes (with a limited detection range) aimed at ground/sky {surveillance} or at street search across two opposite walls. Besides, in the scenario addressed in~\cite{10.1117/12.2325637}, a standard rotating radar (based on LOS) covering the flight space completes the network.
	A network of low-cost Commercial-Off-The-Shelf (COTS) radars, mounted on the facades of buildings or the street lamp sites, {is designed in~\cite{9079047}} to detect  small drones and establish a continuous coverage in urban environment.
	
	{A different perspective is provided in the present study where} RIS technology is proposed to extend the coverage of a standard radar system {whereby the direct LOS from a prospective target is missing}. The basic idea is to place one or more RISs in suitable positions (fixed or possibly deployable on request together with a portable radar) within the operating environment such that there is always a direct path between the radar and each RIS. Besides, in each shadow region (where the direct radar-target path is absent), there is at least a RIS with a LOS toward the target. Thus, the region under test is scanned according to two different radar {operative} modes. The former is for the search in areas with a direct path radar-target and is nothing more than a classic radar {modality}. The latter, designed for operation in N-LOS areas, is based on the formation of a smart and controlled propagation environment where the radar focuses the radiation on a specific RIS. The RIS parameters are suitably set to perform scanning within its area of competence. Hence, after target backscattering and another programmed reflection of the RIS to the radar, a two-way double-{hop} channel is established which allows to accomplish the surveillance task. Alternation between the operative modes ensures the coverage of the entire region of interest.
	
	For the new N-LOS {modality}, the radar equation for {the} Signal to Noise Ratio (SNR) budget~\cite{Skolnik2008}, \cite{barton2013radar}, \cite[ch. 2]{richards1},  is derived including the RIS effects as well as the presence of the two {hops} in the forward and reverse link. 
	Moreover, the Signal to Clutter Ratio (SCR) is computed accounting for either surface	or volume clutter. A discussion on resolution parameters connected with the new mode is provided together with the procedure for data acquisition and formation of the fast-time slow-time data matrix. Remarkably, after this step, {classic and consolidated} range-Doppler processing, possibly {leveraging} clutter cancellation procedures and Constant False Alarm Rate (CFAR) techniques, {can be} used. At the analysis stage {the} SNR and {the} detection Probability ($\text{P}_\text{d}$) are evaluated to assess the performance of the N-LOS mode and to highlight the role of the system parameters with special emphasis on those characterizing the RIS.

	{The paper is organized as follows.	Section II presents some generalities on RIS. Section III introduces the RIS-{assisted} system for radar surveillance in N-LOS environments. In Section IV, the radar equation is reformulated to account for the RIS presence in the computation of the SNR. The expressions of the SCR are derived in Section V. 
		Section VI discusses the use of the radar timeline, data acquisition, and resolution issues for the N-LOS mode, whereas numerical results are presented in Section VII.
		Finally, Section VIII draws some conclusions and highlights possible future research avenues.}
	
	\subsection{Notation}
	Boldface is used for vectors $\ba$ (lower case), and matrices $\bA$ (upper case). The all-ones column vector of size $N$ is indicated as ${\bf 1}_N$. The transpose {and the conjugate} operators are denoted by the symbols $(\cdot)^{\mathrm{T}}$ and {$(\cdot)^*$, respectively. Besides,} the Hadamard (i.e., elementwise) product is indicated as $\odot$.
	The set of $N$-dimensional column vectors of complex numbers is represented by ${\mathbb{C}}^N$ whereas ${\mathbb{C}}^{N,M}$ denotes the set of $N\times M$-dimensional matrices of complex numbers. The letter $j$ indicates the imaginary unit (i.e., $j=\sqrt{-1}$). For any complex number $x$, $|x|$ indicates the modulus of $x$.
	Furthermore, for any $x, y \in \mathbb{R}$, $\max(x, y)$ returns the maximum between the two argument values. Finally, for any $\bx \in \mathbb{C}^N$, $\|\bx\|$ denotes the Euclidean norm.
	
	\section{Generalities on RIS}
	A RIS (also known in the open literature as intelligent reflecting surface, smart reflectarray, large intelligent surface, and so on) is a digitally controllable meta-surface~\cite{Cui2014} composed of a very large number~\cite{8811733} of low-cost passive programmable integrated electronic circuits. These operating units, also referred to as  meta-atoms, are appropriately designed (in terms of geometrical shape, size, orientation, etc.) to make their individual signal response adjustable~\cite{8910627}. Besides, they can be placed either in discrete locations~\cite{6648436, Zhang2018}, or distributed with continuity over the surface~\cite{8319526, 8811733}. 
	Compared to a normal surface, a meta-surface is characterized by alterable reflection and refraction properties (consistent with the generalized Snell’s laws~\cite{Yu333}).
	In particular, the introduction of RIS offers the possibility for an instantaneous change of the incident wave shape by manipulating the scattering particles of the meta-surface (not fixed in the productive process) according to the desired behavior or on the basis of a perception-action model~\cite{9140329}.
	Detailed description of the hardware architectures can be found in~\cite{6648436, 8910627}.

	Characterized by single or multi-layer (typically three) planar structures employing Micro Electro-Mechanical Systems (MEMS) switches, varactor diodes, etc., the RIS can be electronically managed by a general-purpose micro-controller~\cite{7510962} to reflect, refract, or absorb the incident signals with independently controlled amplitude and/or phase shifts~\cite{8466374, 8741198, Renzo2019, 8910627, 9086766, 9140329, 10.1002/adom.202000783, 9215972, 9226135, 9326394}. These extra degrees-of-freedom can be exploited to enhance the capabilities and the efficiency of a wireless  system. In particular, with a proper configuration of the units, the RIS is able to extend the coverage of the transmitter beyond its {natural} LOS, with an additional path toward the target (virtual LOS link)~\cite{9326394}. Besides, the low profile, regular shape and lightweight makes the RIS suitable for an easy installation in the environment~\cite{9328302, 9364358, 9326394}, {including} a deployment into buildings facades, room and factory ceilings, and so on (see also~\cite{8741198}). {Two recent research prototypes are presented in~\cite{246282} and \cite{NTTdocomo}; the former is made of inexpensive antennas whereas the latter is a meta-surface.}
	
	{Unlike} phased arrays and relays, RISs {(being passive or nearly passive~\cite{9140329})} do not require the use of power amplifiers and, even if they demand the largest number of cell elements among the mentioned technologies, they can be realized with the least expensive components~\cite{best_readings_in_ris}.
	This is due to the fact that a RIS can be realized via large matrices of inexpensive antennas or metamaterial surfaces (possibly conformal) whose cells have sizes and {inter-element space smaller than the wavelength}.
	Additional observations about the distinctive peculiarities of RIS, as compared with competing technologies, are discusses in~\cite{9140329}.

	\section{RIS-based Radar Operation for N-LOS Surveillance}	
\begin{figure}[h!] 
		\centering
		\includegraphics[width=0.95\linewidth]{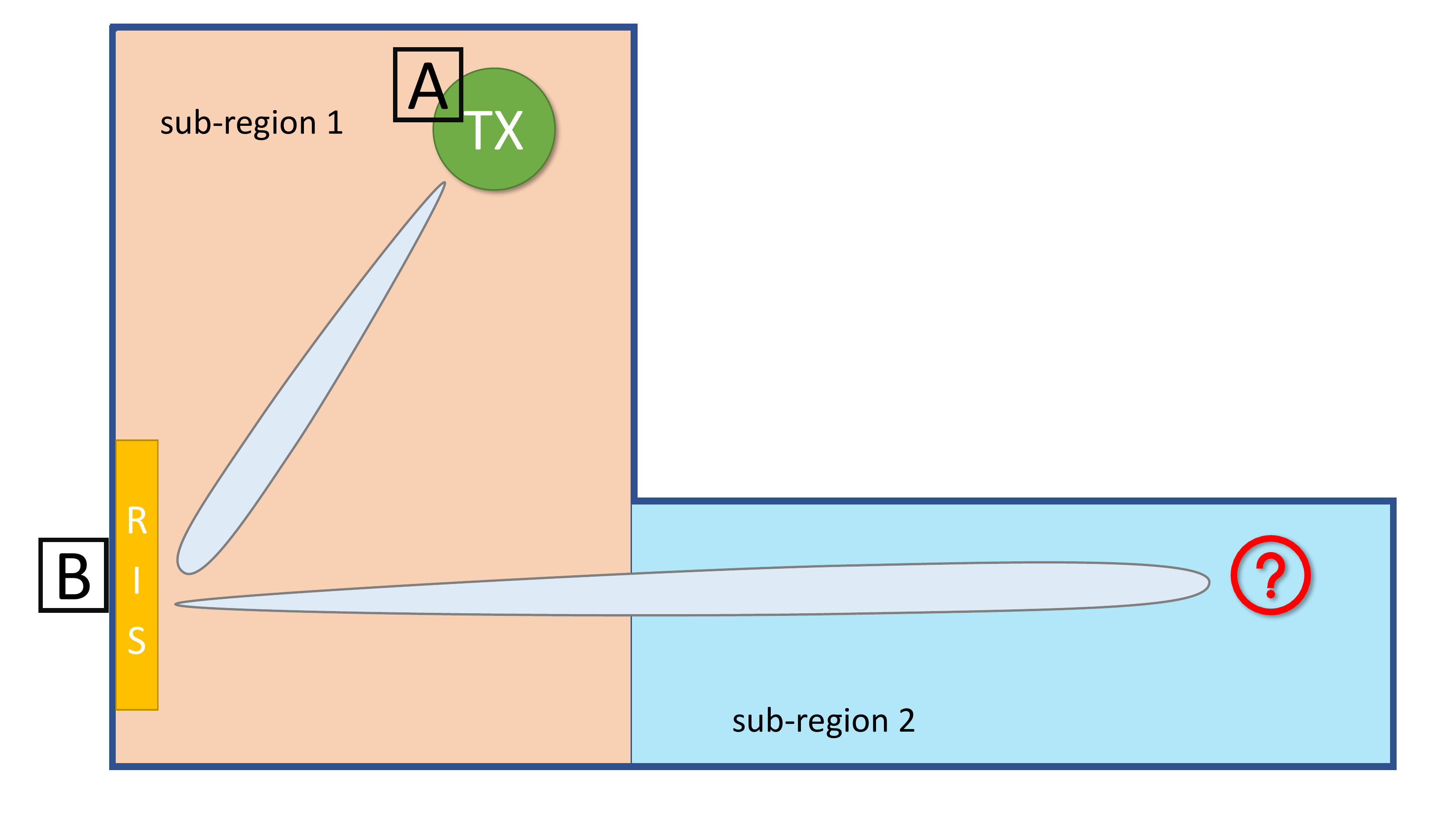}
		\caption{Geometry of the RIS radar system. The prospective target is located in sub-region 2, which is not directly covered by the radar.}
		\label{fig:1} 
	\end{figure}

	This section discusses the use of RISs for radar surveillance without LOS. The idea is to control artificially the propagation environment to extend the radar coverage in those regions which cannot be reached by the direct path. {In this respect, the intelligent surface	plays the role of a	passive but controllable element of the radar environment	which aids the establishment of a radar-target propagation path.} An illustration of the basic concept is now provided. The radar is located at point A and has to cover the shaded region (union of sub-region 1 and sub-region 2) displayed in Fig.~\ref{fig:1}.

\begin{figure}[h!] 
		\centering
		\includegraphics[width=0.95\linewidth]{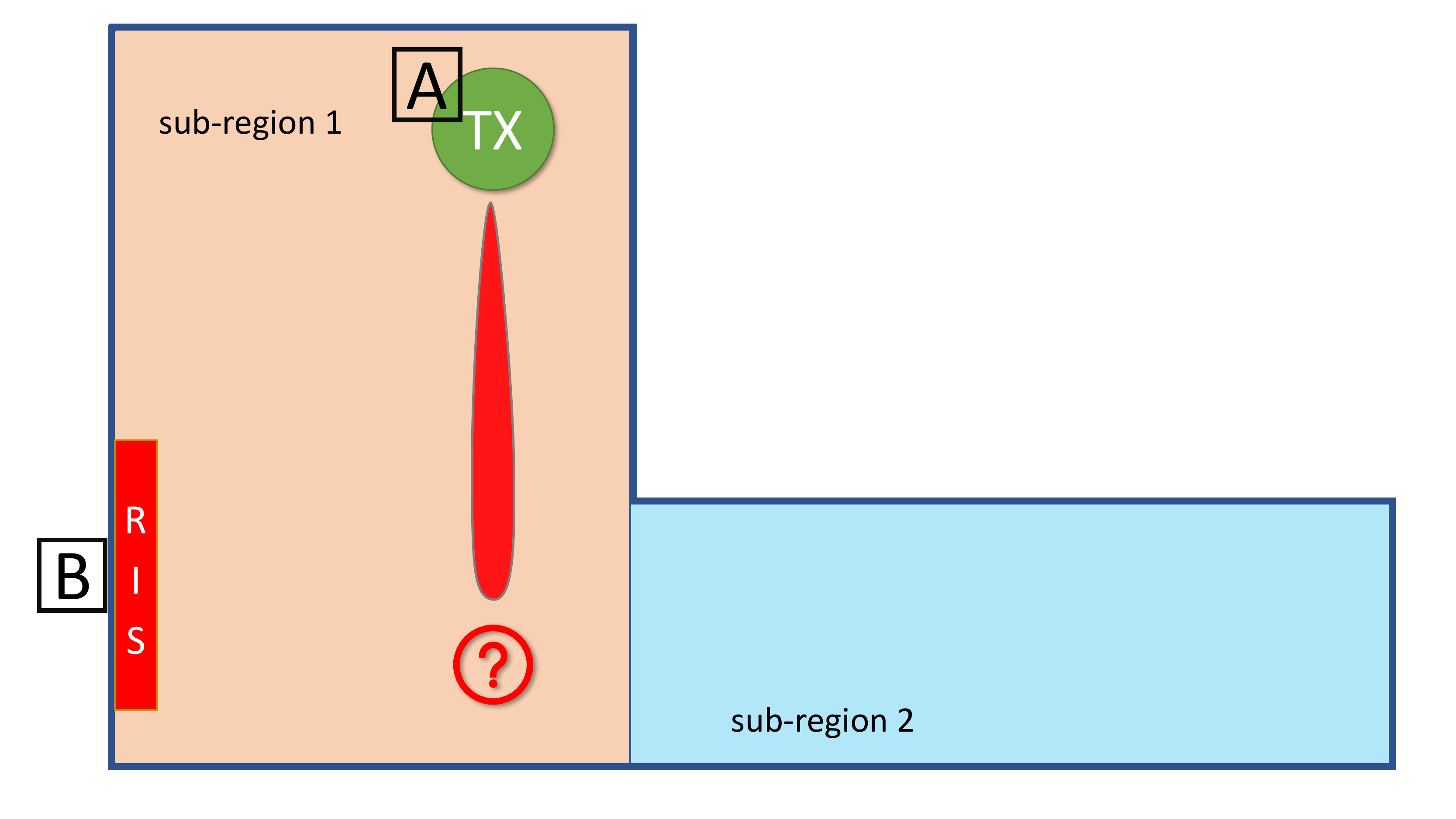}
		\caption{Geometry of the RIS radar system. The prospective target is directly covered by the radar. The RIS is inactive.}
		\label{fig:2} 
	\end{figure}

	While in sub-region 1 a direct path between the radar and the prospective target can be established, in sub-region 2 a LOS path is no longer present. This situation is representative of a typical radar operation in urban environments or in the presence of natural obstacles (for instance hills, {buildings,} or natural/artificial wedges) which actually inhibit the direct path. With a standard radar mode {a} partial coverage of sub-region 2 can be achieved only via natural diffraction or multipath propagation. The approach discussed here (to handle the aforementioned scenarios) relies on the development of innovative radar modes exploiting RISs with the endeavor to build a smart propagation environment capable of ensuring the desired coverage in a controlled way. With reference to the example in Fig.~\ref{fig:1} (the concept can be also generalized to include multiple RISs), a suitable radar transmission-reception protocol leveraging the availability of a RIS in position B can be established.
	Precisely, when the radar performs the standard surveillance operations in sub-region 1, see Fig.~\ref{fig:2},  the RIS is inactive and classic scanning patterns (i.e., sequential scanning, stacked beams, etc.) can be adopted.
	
\begin{figure*}[h!] 
		\centering
		\includegraphics[trim=0 210 170 0,clip,width=0.7\linewidth]{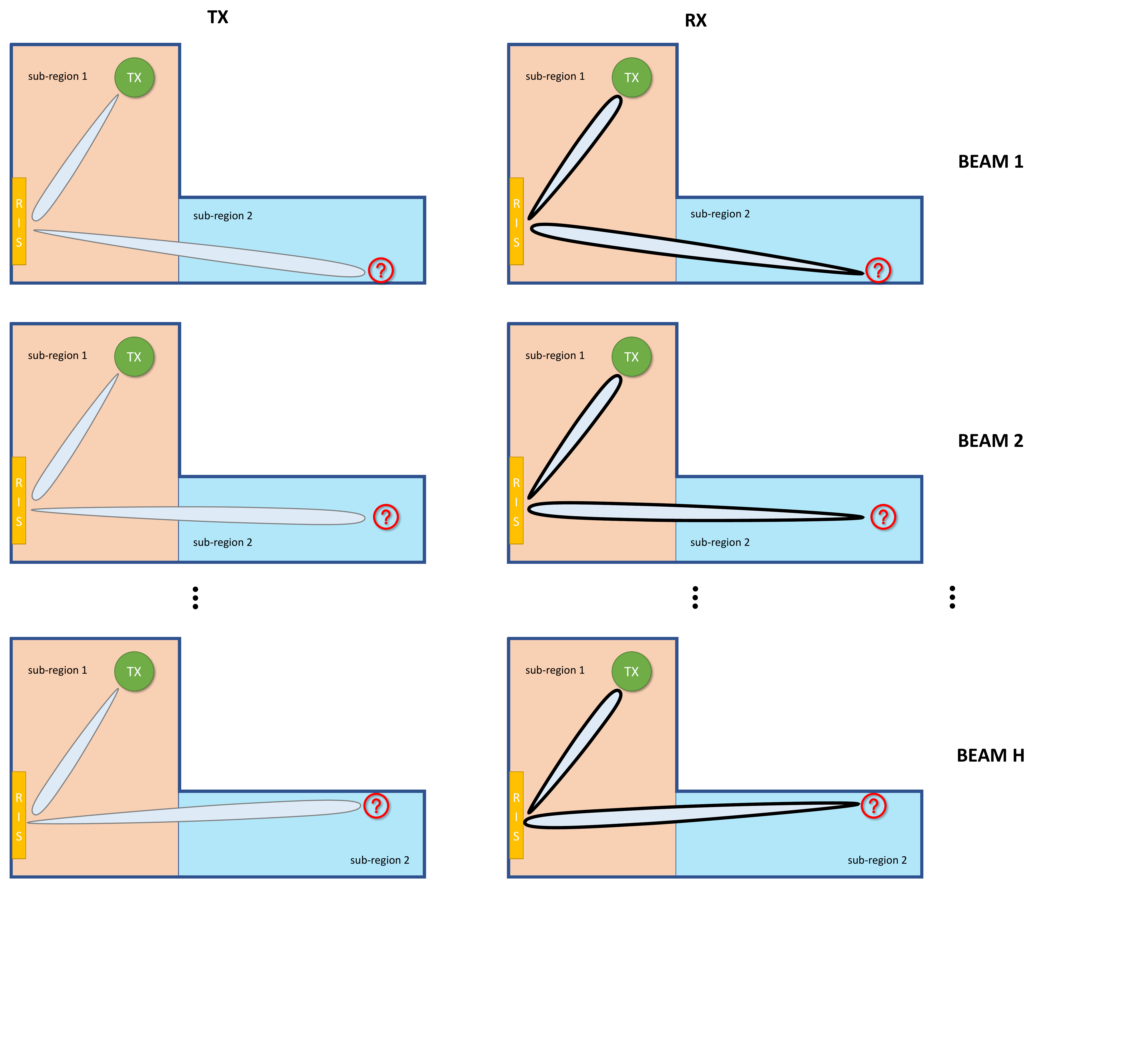}
		\caption{An example of a sequential scanning of sub-region 2. In the $i$-th dwell, the RIS is properly configured to cover the $i$-th angular position, for $i=1, \dots, H$.}
		\label{fig:3} 
	\end{figure*}

	Periodically, according to a desired temporal scheduling, a radar scan in sub-region 2 is demanded. During these time intervals the RIS is turned on and the radar beam is pointed toward the RIS {allowing for the angle} of the reflected wave to be arbitrarily controlled. This can be done adjusting the RIS characteristic parameters for instance through a pre-canned program or a communication-oriented software which represents another degree of freedom (but also a challenge) in the design of the radar scheduler. Resorting to a specific timing procedure, based on the knowledge of the radar and RIS positions the system can scan through sub-region 2. Just to give an example (even if it is worth mentioning that alternative sweep patterns can be conceived) in Fig.~\ref{fig:3}, a sequential scanning protocol is displayed where in a specific dwell time transmission and reception are alternated.

	Let us focus on the $i$-th dwell and describe the sequence of involved operations. The radar focuses the beam on the RIS which reflects the radiation to cover the $i$-th angular sector of sub-region 2. Then the listening time starts where the RIS is programmed to steer the target radiation toward the radar. Hence, the procedure is repeated for the $N_p$ pulses of the dwell. All the signal acquisition as well as signal and data processing is performed at the radar side also accounting for the extra-path (two-way) between the radar and RIS. Then the scanning of sub-region 2 continues in order to cover the entire area of interest according to different beams (see beams 1, ..., H in Fig.~\ref{fig:3}).     
	
	\section{Radar Equation for Sub-region 2}\label{radar_eq_2_sec}
	
	While the standard radar equation can be used to describe the coverage in sub-region 1, a suitable reformulation is necessary for the operation in sub-region 2, to explicitly account for the effects of the RIS  and obtain a design tool for system sizing.
	
\begin{figure}[h!] 
		\centering
		\includegraphics[width=0.95\linewidth]{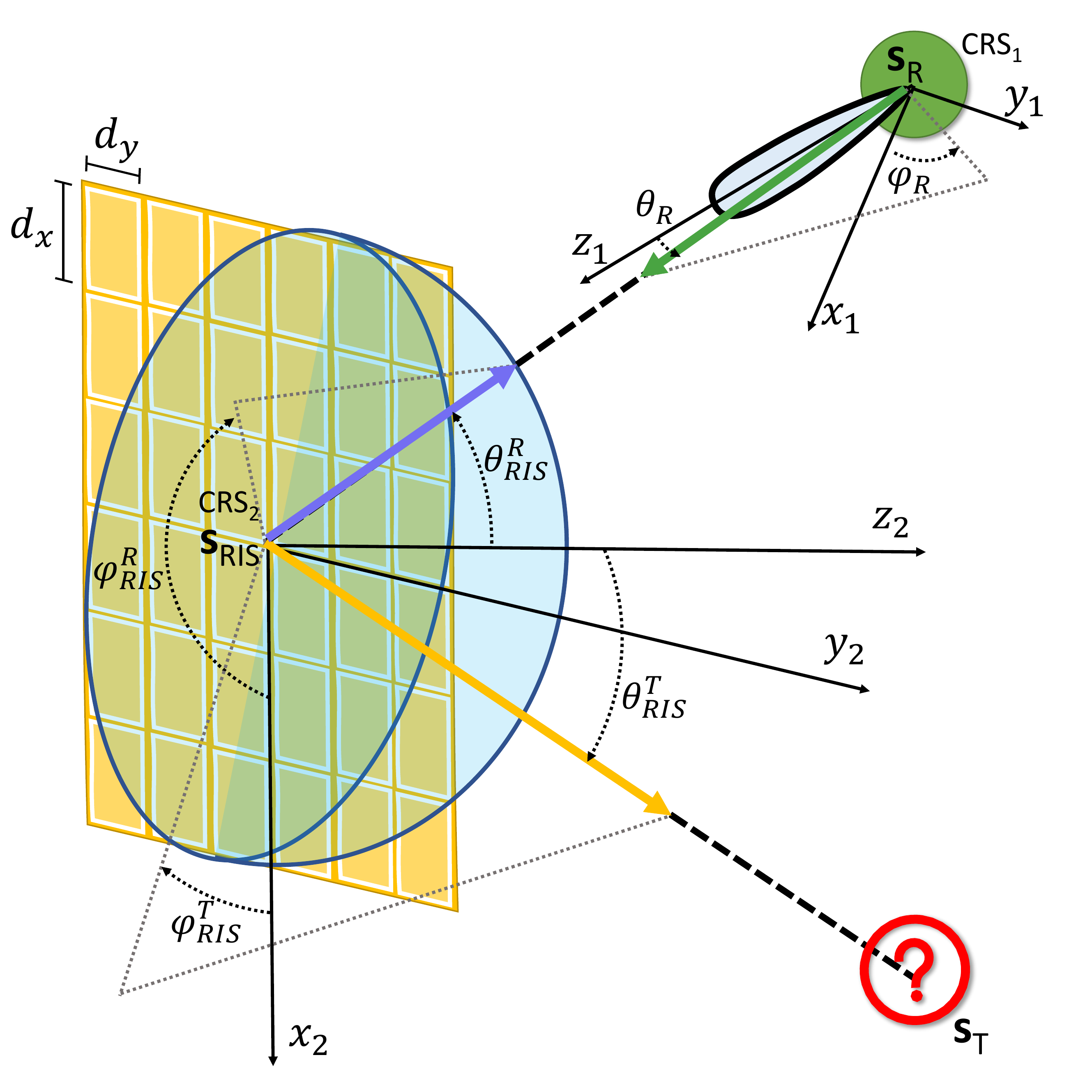}
		\caption{Model of the RIS-aided sensing system with both the radar and the target assumed located in the far field region of the RIS and vice-versa.}
		\label{fig:4} 
	\end{figure}

	In the following it is supposed that the RIS is in the far-field region of both the radar and the target.
	To proceed further, let us denote by $\bs_R$, $\bs_{RIS}$, and $\bs_T$ the phase-center position of the radar antenna, the RIS, and the target respectively, see Fig.~\ref{fig:4}. Therein, two Cartesian Reference Systems (CRSs), referred to hereafter as $\mbox{CRS}_1$ and $\mbox{CRS}_2$, are reported. The former is centered at $\bs_R$ and its $z$-axis coincides with radar antenna pointing direction, the latter is centered at $\bs_{RIS}$ with the $z$-axis steered toward the  normal to the RIS. Besides,  $\phi_R$ and $\theta_R$ refer to the azimuth and elevation angles of the RIS  with respect to (w.r.t.) $\mbox{CRS}_1$, respectively, whereas $\phi_{RIS}^R$ and $\theta_{RIS}^R$ indicate the azimuth and elevation  angles of the radar w.r.t. $\mbox{CRS}_2$. Furthermore, $\phi_{RIS}^{T_a}$  and $\theta_{RIS}^{T_a}$ represent the azimuth and elevation angles of the target w.r.t. $\mbox{CRS}_2$. Finally, $C_{i,h}$, $i=1,\ldots,N$, $h=1,\ldots,M$, denotes the unit cell/patch of the RIS whose center is located, w.r.t. $\mbox{CRS}_2$, at $\bs_{i,h}=\left[(-\frac{(N-1)}{2}+i-1) d_x, (-\frac{(M-1)}{2}+h-1) d_y, 0 \right]^T$, where, without loss of generality, it is assumed that both $N$ and $M$ are odd numbers.  
	
	Let us now focus on the derivation of the radar-range equation for a RIS-assisted surveillance system and denote by
	\begin{itemize}
		\item $P_T$ the radar peak power;
		\item $G_T$ the radar transmit antenna gain;
		\item $r_1=\|\bs_R-\bs_{RIS}\|$ the distance between the radar and the RIS;
		\item $F^R(\bar{\theta},\bar{\phi})$ the  normalized radar power radiation pattern in the look direction $(\bar{\theta},\bar{\phi})$ w.r.t. $\mbox{CRS}_1$.
	\end{itemize}
	Hence, indicating by $\bE^{in}$ the matrix collecting in the $(i,h)$-th entry the electric field {(in a given polarization)} impinging on the RIS-patch $C_{i,h}$ (due to the radar illumination), it follows, under the narrow-band assumption, that 
	\begin{eqnarray}
		\bE^{in}={\sqrt{2 Z_0 {\cal{P}}^{in}}}\bS_1 e^{j \phi_1} ,
	\end{eqnarray}
	where
	\begin{itemize}
		\item $
		{\cal{P}}^{in}=\frac{G_T P_T F^R(\theta_R,\phi_R)}{4 \pi r_1^2}$ is the plane-wave spatial power density at the RIS;
		\item $\bS_1 = \bp_1(\theta_{RIS}^R,\phi_{RIS}^R)\bp_2(\theta_{RIS}^R,\phi_{RIS}^R)^T \in \mathbb{C}^{N,M}$  is the RIS steering matrix {associated with radar wave direction of arrival} (accounting for the electromagnetic path-lengths between the radar phase-center and the different RIS-patches), with $$\bp_1(\theta_{RIS},\phi_{RIS})\!\!=\!\![e^{j \pi \frac{d_x}{\lambda_0} (1\!-\!N) u_{RIS}}\!,\ldots,\! e^{j\pi \frac{d_x}{\lambda_0} (N\!-\!1) u_{RIS}}]^T\!\!,$$ 
		$$\bp_2(\theta_{RIS},\phi_{RIS})\!\!=\!\![e^{j\pi \frac{d_y}{\lambda_0} (1\!-\!M) v_{RIS} }\!,\ldots,\!e^{j\pi \frac{d_y}{\lambda_0} (M\!-\!1) v_{RIS}}]^T\!\!,$$ 
		{the vertical and horizontal RIS manifold vectors,} where $u_{RIS}=\sin(\theta_{RIS})\cos(\phi_{RIS})$, $v_{RIS}=\sin(\theta_{RIS})\sin(\phi_{RIS})$ {are the directional cosines} and $\lambda_0$ is the radar operating wavelength;
		\item $Z_0$ is the characteristic impedance of the medium;
		\item  $\phi_1$ accounts for  the phase term related to the propagation path between radar and RIS as well as the phase response of the radar transmit antenna. 
	\end{itemize}
	Now, denoting by
	\begin{itemize}
		\item $F(\theta,\phi)$ the normalized RIS-patch power radiation pattern at the look direction $(\theta,\phi)$ w.r.t. {$\mbox{CRS}_2$},
		\item $\eta_{RIS}$ the unit patch efficiency  (assumed, for simplicity, common to all the RIS-patches), which accounts for taper and spillover effects~\cite{9265248},
	\end{itemize}
	the power gathered by any RIS-patch (due to the radar illumination) is
	\begin{eqnarray}\label{eq_1}
		P^{RIS}&=&\frac{|\bE(i,h)|^2}{2 Z_0}F(\theta_{RIS}^R,\phi_{RIS}^R)d_x d_y \eta_{RIS}\nonumber\\
		&=&\frac{G_T P_T}{4 \pi r_1^2}F^R(\theta_R,\phi_R)F(\theta_{RIS}^R,\phi_{RIS}^R)d_x d_y \eta_{RIS}\nonumber ,
	\end{eqnarray}
	where $F(\theta_{RIS}^R,\phi_{RIS}^R)d_x d_y \eta_{RIS}$ is effective antenna area of any RIS-patch. Hence, the electric field {\em artificially reflected} by $C_{i,h}$ and impinging on the target is given by the $(i,h)$-th entry of the matrix 
	\begin{eqnarray}\label{eq_3}
		\bE^{Target}= \sqrt{\frac{2 Z_0 G P^{RIS} F(\theta_{RIS}^{T_a},\phi_{RIS}^{T_a})}{4 \pi r_2^2} } \: \bSigma \: e^{j\phi_1} e^{j\phi_2} ,
	\end{eqnarray}
	where 
	\begin{itemize}
		\item $\bSigma = \bS_1\odot \bGamma\odot \bS_2$;
		\item $\bS_2=\bp_1(\theta_{RIS}^{T_a},\phi_{RIS}^{T_a})\bp_2(\theta_{RIS}^{T_a},\phi_{RIS}^{T_a})^T$,
		\item $\bGamma\in \mathbb{C}^{N,M}$  contains at the $(i,h)$-th entry  the programmable reflection coefficients associated with $C_{i,h}$,
		\item $G$ is the antenna power gain of a RIS-patch;
		\item $r_2=\|\bs_T-\bs_{RIS}\|$; 
		\item $\phi_2$ is the phase term related to $r_2$. 
	\end{itemize}
	As a result, the electromagnetic power density {impinging on the target} is
	\begin{equation}\label{eq_4}
		{\cal{P}} \;=\; {\frac{G_T P_T G F^{tot} d_x d_y \eta_{RIS}}{16 \pi^2 r_1^2 r_2^2  } }|{\bf 1}_N^T \: \bSigma \: {\bf 1}_M|^2 
	\end{equation}
	with $F^{tot}=F^R(\theta_R,\phi_R)F(\theta_{RIS}^R,\phi_{RIS}^R)F(\theta_{RIS}^{T_a},\phi_{RIS}^{T_a})$, where to ease the notation the explicit dependence over the angles has been omitted.
	
	Letting $\sigma$ the monostatic RCS of the target along the {LOS} with the RIS, the power reflected back by the target toward the RIS is,	$$P_{refl}={\cal{P}}\sigma.$$ 
	Hence, supposing that the RIS does not change its programmable reflecting coefficients in the backward path and the radar antenna pattern is the same at the transmission and reception stages, i.e., reciprocity holds true, it immediately follows that the radar received power is
	\begin{eqnarray}\label{eq_5}
		\begin{aligned}
			{{P}^{rx}}= {\frac{G_T^2 P_T G^2 {F^{tot}}^2 d_x^2 d_y^2 \eta_{RIS}^2\lambda_0^2 \sigma}{r_1^4 r_2^4  (4\pi)^5} }|{\bf 1}_N^T \: \bSigma \: {\bf 1}_M|^4
		\end{aligned}
	\end{eqnarray}
	
	Leveraging (\ref{eq_5}), it is now possible  to derive the SNR for a RIS-assisted sensing mode. To this end, let $F_N$ be the noise figure of the radar receiver and $B$ the radar bandwidth, then
	\begin{eqnarray}\label{eq_6}
		\begin{aligned}
			{\mbox{SNR}} =& {\frac{G_T^2 P_T G^2 {F^{tot}}^2 d_x^2 d_y^2 \eta_{RIS}^2\lambda_0^2 \sigma}{r_1^4 r_2^4  (4\pi)^5 k T_0 B F_N} }|{\bf 1}_N^T \: \bSigma \: {\bf 1}_M|^4 = \\
			& {\frac{G_T^2 P_T G^2 {F^{tot}}^2 d_x^2 d_y^2 \eta_{RIS}^2\lambda_0^2 \tau \sigma}{r_1^4 r_2^4  (4\pi)^5 k T_0 F_N} }|{\bf 1}_N^T \: \bSigma \: {\bf 1}_M|^4 ,
		\end{aligned}
	\end{eqnarray}
	where $k$  is Boltzmann’s constant,  $T_0$ the standard temperature, i.e., 290 K,  and $\tau$ the pulse length. Remarkably, the provided SNR expression in terms of  the pulse length $\tau$ holds true for both unmodulated and modulated pulses.
	
	Usually, a radar system transmits multiple pulses to probe the environment. Equation (\ref{eq_6}) can be used to determine the integrated SNR when a coherent burst of $N_p$ pulses illuminates the target via a RIS. Specifically, if the RIS does not change its reflecting characteristics during the radar dwell-time, the RCS exhibits at most a burst-to-burst fluctuation, and  the $N_p$ echoes are combined through coherent integration processing, the resulting coherently integrated SNR is given by
	\begin{eqnarray}\label{eq_7}
		{\mbox{SNR}}_c(N_p)= {\mbox{SNR}} N_p ,
	\end{eqnarray}
	where $N_p$ is the coherent integration gain.
	
	Expression (\ref{eq_7}) represents an idealized form of the target SNR, because phenomena that reduce the received useful signal power,  producing SNR loss, are neglected. Otherwise stated, a more realistic SNR expression is
	\begin{eqnarray}\label{eq_8}
		{\mbox{SNR}}_c(N_p)= \frac{\mbox{SNR} N_p}{L_{s}} ,
	\end{eqnarray}
	where $L_s$ {constitutes} the total system loss. This last term can be also recast as
	$$L_s=L_t L_{atm} L_r L_{sp} L_{ris}$$
	with
	\begin{itemize}
		\item $L_t$ the transmit loss: it is related to the signal attenuation due to transmitter devices and components, such as circulators and waveguides;
		\item $L_{atm}=L_{atm_1} L_{atm_2}$ the atmospheric loss: it accounts for electromagnetic wave absorption due to fog, rain, snow, and water vapor to mention a few ($L_{atm_i}$, $i=1,2$,  refers to the two-way propagation loss in sub-region $i$);
		\item $L_r$ the receiver loss: it represents the receive-side counterpart to $L_t$;
		\item $L_{sp}$ the signal processing loss: it accounts for the coherent integration gain reduction due to range/Doppler straddle loss, and useful signal mismatches ({e.g. from possible I/Q imbalance} and oscillators phase noise); besides, it describes sub-optimal processing, e.g., CFAR detectors and range/Doppler tapering;
		\item $L_{ris}$ the RIS loss: it refers to RIS-patches power absorption as well as mismatches between nominal and actual reflecting coefficients, resulting from quantization errors. 
	\end{itemize}
	Based on {eqs. (\ref{eq_6}) and (\ref{eq_8})}, it is immediate to obtain the average power form of the radar range equation for a RIS-assisted sensing mode. In this respect, note that
	$$P_T=P_{avg}\frac{T}{\tau},$$
	where $T$ is the Pulse Repetition Interval (PRI) and $P_{avg}$ the average transmit power. Hence, denoting by $T_d=N_p T$ the radar dwell time, it follows that (\ref{eq_8}) can be cast as 
	\begin{equation}\label{eq_SNR_c}
		{\mbox{SNR}}_c= {\frac{G_T^2  G^2 {F^{tot}}^2 d_x^2 d_y^2 \eta_{RIS}^2\lambda_0^2 \sigma P_{avg} T_d}{r_1^4 r_2^4  (4\pi)^5 k T_0 F_N L_s} }|{\bf 1}_N^T \: \bSigma \: {\bf 1}_M|^4 ,
	\end{equation}
	which clearly highlights that the longer the dwell time the higher the SNR, as per a conventional radar system.
	
	Before concluding this section, {two} interesting remarks are provided.\\
	{\bf Remark 1.} The SNR depends on $|{\bf 1}_N^T \; \bSigma \; {\bf 1}_M|^4$, in the following referred to as RIS-induced pattern, which is a function of the RIS angular location w.r.t. the radar, i.e., $\left(\theta_{RIS}^R,\phi_{RIS}^R\right)$, the target angular location w.r.t. RIS, i.e., $\left(\theta_{RIS}^{T_a},\phi_{RIS}^{T_a}\right)$, and the RIS programmable reflecting coefficients, i.e., $\bGamma$. The RIS-induced pattern is upper bounded by $(NM)^4$ because the  amplitude of each entry of $\bGamma$ is {assumed smaller than} or equal to one (any RIS-patch is {modeled as} a passive component) and the entries of $\bS_i$, for $i=1,2$, have a constant modulus. The mentioned bound is achieved when $\bGamma={\bS_1}^*\odot {\bS_2}^*$, which is tantamount to coherently aligning the phases of {all the contributions reflected from the RIS}, by compensating the phases terms of ${\bS_1}\odot {\bS_2}$. However, while $\left(\theta_{RIS}^R,\phi_{RIS}^R\right)$ is fixed and known with a high accuracy, only a nominal value of $\left(\theta_{RIS}^{T_a},\phi_{RIS}^{T_a}\right)$ { is available, which coincides} with the steering direction of the RIS beam in sub-region 2. This can
	determine a mismatch between the target actual angular position and the RIS steering direction which can lead 
	to a SNR loss. To shed light on the effects of the mentioned mismatches and to provide as a by-product, guidelines for the definition of a sequential scanning protocol in sub-region 2 {(in particular how to space the pointing directions)}, let us assume that $\bGamma=\bar{\bS_1}^*\odot \bar{\bS_2}^*$ with
	$\bar{\bS_1}=\bp_1(\theta_1,\phi_1)\bp_2(\theta_1,\phi_1)^T$ and $\bar{\bS_2}=\bp_1(\theta_2,\phi_2)\bp_2(\theta_2,\phi_2)^T$. As shown in Appendix \ref{RIS-ind}, the RIS-induced pattern boils down to
	\begin{eqnarray}\label{patter_exp}
		|{\bf 1}_N^T \: \bSigma \: {\bf 1}_M|^4\!\!\!\!&=&\!\!\!\!\!\left|\!\frac{\sin(\Delta \kappa_u\pi N)}{\sin(\Delta \kappa_u\pi)}\!\right|^4\left|\!\frac{\sin(\Delta \kappa_v\pi M)}{\sin(\Delta \kappa_v\pi)}\!\right|^4  
	\end{eqnarray}
	where
	$$\Delta \kappa_u=\frac{d_x}{\lambda_0}\left[(u_{RIS}^{T_a}-u_2)+(u_{RIS}^R-u_1)\right],$$
	and
	$$\Delta \kappa_v=\frac{d_y}{\lambda_0}\left[(v_{RIS}^{T_a}-v_2)+(v_{RIS}^R-v_1)\right],$$
	with
	$$u_i=\sin(\theta_i)\cos(\phi_i),\,v_i=\sin(\theta_i)\sin(\phi_i),\,i=1,2,$$
	and, for $\beta \in\{R,T_a\}$,
	$$u_{RIS}^\beta=\sin(\theta_{RIS}^{\beta})\cos(\phi_{RIS}^{\beta}), v_{RIS}^{\beta}=\sin(\theta_{RIS}^{\beta})\sin(\phi_{RIS}^{\beta}).$$
	
	{Equation~\eqref{patter_exp} suggests that} the RIS-induced pattern exhibits a factorized form w.r.t. the variables $\Delta \kappa_u$ and $\Delta \kappa_v$, which depends on {the offsets of the} directional cosines. Specifically, $\Delta \kappa_u$ is the sum of the offsets in the $u$-direction $(u_{RIS}^R-u_2)$ and $(u_{RIS}^T-u_1)$, associated with the radar-RIS path and the RIS-target path, respectively, times $\frac{d_y}{\lambda_0}$. Analogously,  $\Delta \kappa_v$ accounts for the offsets in the $v$-direction.
	
	Each factor is the fourth power of a specific Dirichlet kernel. Precisely, the pattern in the $\Delta \kappa_u$ domain presents a maximum at {$\Delta \kappa_u=0$}, with a $\frac{2}{N}$ null-to-null mainlobe width. Besides, in each unitary-period, it exhibits $N-2$ sidelobes whose width is $\frac{1}{N}$. Similarly, in the {$\Delta \kappa_v$} domain, the mainlobe is centered at {$\Delta \kappa_v=0$} with a $\frac{2}{M}$ width, and $M-2$ {sidelobes} of width $\frac{1}{M}$ appear in each unitary-period.
	
	Based on the above considerations, it follows that the maximum {value} of the RIS-induced pattern {is appearing in} $\theta_{RIS}^{R}= \theta_1$, $\phi_{RIS}^{R}= \phi_1$, $\theta_{RIS}^{T_a}= \theta_2$, and $\phi_{RIS}^{T_a}= \phi_2$. Let us now suppose that the phase-matching condition holds true only in sub-region 1. Then, (\ref{patter_exp}) implies that the RIS-induced pattern, in sub-region 2, is the same as for a uniform rectangular array disposed in the RIS plane (with $N$ and $M$ antennas along the $x$-axis and $y$-axis in $\mbox{CRS}_2$, respectively) which is electronically steered in the direction $\theta_2$ and $\phi_2$. As a consequence, the  azimuth and elevation single-side beamwidths $\bar{\phi}_{RIS}$ and $\bar{\theta}_{RIS} $ can be obtained as in \cite{vantrees4}; for instance, if $d_x=d_y=\frac{\lambda_0}{2}$ and $\theta_2=0$, i.e., the RIS-patches are steered in the broadside direction, with {$\bar{\phi}_{RIS}=0.891/N$} and {$\bar{\theta}_{RIS}=0.891/M$} the $12$ dB beamwidths of the RIS. 
	
	Finally, it is worth pointing out that eq. (\ref{patter_exp}) provides a guideline to tile the coverage area in sub-region 2 with RIS beams (i.e., appropriate selections of $\theta_2$ and $\phi_2$), also accounting for the maximum acceptable SNR angular loss and the dwell time, {see~\eqref{eq_SNR_c}}.
	
	{\bf Remark 2.} An interesting interpretation of {eqs.} (\ref{eq_6}) and (\ref{eq_8}) {is now provided in terms of an {\em equivalent} monostatic radar configuration}. Specifically, eq. (\ref{eq_8}) can be cast as
	\begin{eqnarray}\label{eq_10}
		{\mbox{SNR}_{c}(N_p)}\!\!&=&\!\! {\frac{{G_T^{eq}}^2 P_T \lambda_0^2 {\sigma^{eq}} N_p \tau}{(r_1+r_2)^4  (4\pi)^3 k T_0 F_N L_s^{eq}} } \:,
	\end{eqnarray}
	where
	$$\sigma^{eq}=\sigma,\, G_T^{eq}=G_T G \, \eta_{RIS}\, N^2 M^2,$$ 
	$(r_1+r_2)$ is the {\em total} target range, and
	$$L_s^{eq}=L_t L_{atm} L_r L_{sp}^{eq} L_{ris}^{eq} L_{geom} ,$$ with
	\begin{eqnarray}
		L_{sp}^{eq}&=& L_{sp} \frac{(NM)^4}{{F^{tot}}^2|{\bf 1}_N^T \: \bSigma \: {\bf 1}_M|^4}, \nonumber\\ 
		L_{ris}^{eq}&=&{L_{ris}\pi^2},\nonumber\\
		L_{geom}&=&\left(\frac{2}{\frac{\sqrt{d_x d_y}}{r_1}+\frac{\sqrt{d_x d_y}}{r_2}}\right)^4.\nonumber
	\end{eqnarray}
	Interestingly, $L_{geom}$ can be also expressed as 
	$$L_{geom}=\left(\frac{\left( \frac{r_1^{-1}+r_2^{-1}}{2}\right)^{-1}}{\sqrt{d_x d_y}}\right)^4,$$
	which coincides with the fourth power of the harmonic mean between the path lengths in the two sub-regions, i.e., $(1/2\:(1/r1 + 1/r2))^{-1}$, normalized to the geometric mean between $d_x$ and $d_y$, i.e., $\sqrt{d_x d_y}$, which represents the characteristic size of each RIS-patch. 
	As consequence, $L_{geom}$ accounts for the loss {induced by} a different propagation mechanism that relies on the reflection {of the waves} from RIS-patches, each with area $d_x d_y$.
	
\begin{figure}[h!] 
		\centering
		\includegraphics[width=0.95\linewidth]{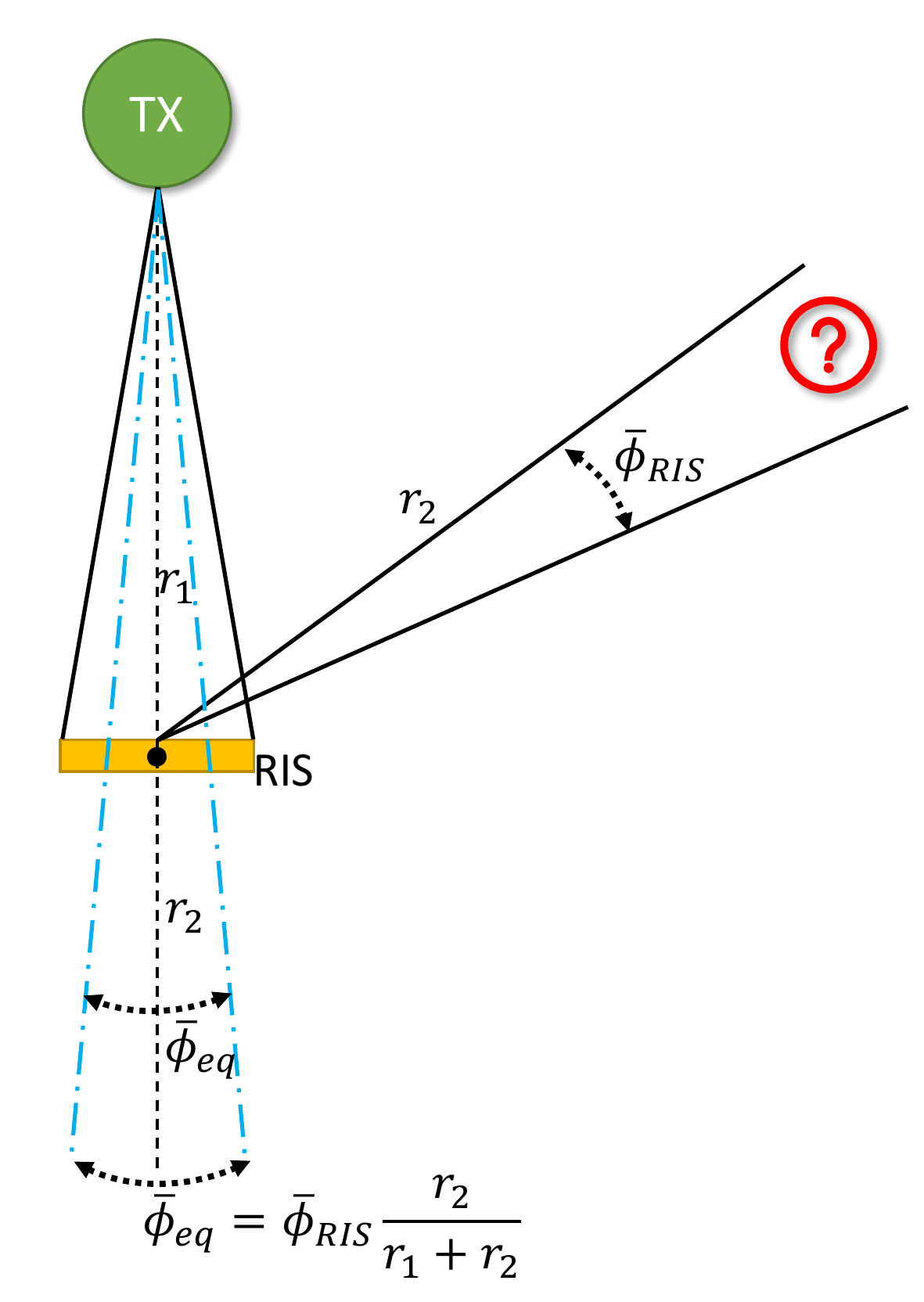}
		\caption{A representation of the {\em equivalent} monostatic configuration associated with the RIS-assisted surveillance system.}
		\label{fig:6} 
	\end{figure}

	A pictorial representation of the {\em equivalent} monostatic configuration associated with the RIS-assisted sensing system is depicted in Fig.~\ref{fig:6}. 
	{Therein, a 2D situation is reported and the azimuth beamwidth of the equivalent monostatic system, i.e. $\bar{\phi}_{eq}$, is set so that the cross-range resolution at the equivalent point of interest, i.e., $\bar{\phi}_{eq} (r_1+r_2)$, is the same as that of the RIS-assisted system at the point $\bs_T$, i.e., $\bar{\phi}_{RIS} \: r_2$.}

	\section{Radar Equation for Clutter Interference}
	This section is devoted to the computation of the SCR for the operative mode in sub-region 2. Surface clutter and volume clutter are considered. In the first case, the SCR is evaluated for pulse length-limited and beamwidth-limited geometry. 
	\subsection{Surface clutter: pulse length-limited}
	The illuminated area $A_c$ depends on the RIS azimuth beamwidth $\bar{\phi}_{RIS}$ and the length of the pulse $\tau$ measured along the surface; specifically,
	\begin{equation}
		\begin{aligned}
			A_c & = 2 \: r_2 \left(\frac{c \tau}{2}\right) \tan{\left(\frac{\bar{\phi}_{RIS}}{2}\right)} \sec{\psi} \\
			& \simeq r_2\: \left(\frac{c \tau}{2}\right) \bar{\phi}_{RIS}  \sec{\psi} ,\nonumber
		\end{aligned}	
	\end{equation}
	where $\psi$ is the grazing angle and the {tangent} approximation relies on the small angle assumption for $\bar{\phi}_{RIS}$. Letting $\sigma_0$ the surface reflectivity, the clutter power at the radar is thus given by
	\begin{eqnarray}
		\begin{aligned}
			&C \; = \; {\frac{G_T^2 P_T G^2 {F^{tot}}^2 d_x^2 d_y^2 \eta_{RIS}^2\lambda_0^2  A_c \sigma_0}{r_1^4 r_2^4  (4\pi)^5} }|{\bf 1}_N^T \: \bSigma \: {\bf 1}_M|^4 =  \\
			&  {\frac{G_T^2 P_T G^2 {F^{tot}}^2 d_x^2 d_y^2 \eta_{RIS}^2\lambda_0^2 c \tau \bar{\phi}_{RIS} \: \sigma_0 \: \sec{\psi}}{r_1^4 r_2^3 2^{11} \pi^5} }|{\bf 1}_N^T \: \bSigma \: {\bf 1}_M|^4 . \nonumber
		\end{aligned}
	\end{eqnarray}
	It follows that the SCR is given by 
	\begin{equation}
		{\mbox{SCR}}_{pll} = \frac{2 \: \cos{\psi}}{r_2 \:c\: \tau \: \bar{\phi}_{RIS}} \frac{\sigma}{\sigma_0} ,
	\end{equation}
	which highlights that the SCR decreases according to an inverse linear law with the {$r_2$} distance between the RIS and the target. 
	{Besides, the SCR does not depend on the $r_1$ distance between the radar and the RIS, nor on the radar transmit/receive beamwidths. Moreover, the SCR is inversely} proportional to RIS azimuth beamwidth $\bar{\phi}_{RIS}$, which {frames} the size of the radar cell in sub-region 2.

	\subsection{Surface clutter: beamwidth-limited}
	Assuming a circular beam, the illuminated area is
	\begin{equation}
		A = \pi \: r_2^2 \: \tan^2{\left(\frac{\bar{\phi}_{RIS}}{2}\right)} \simeq \frac{\pi \: r_2^2 \: \bar{\phi}_{RIS}^2}{4} ,
	\end{equation}
	where again the small angle approximation for $\bar{\phi}_{RIS}$ has been invoked. 
	As a consequence, the clutter power at the radar is
	\begin{eqnarray}
		\begin{aligned}
			C= \frac{G_T^2 P_T G^2 {F^{tot}}^2 d_x^2 d_y^2 \eta_{RIS}^2\lambda_0^2 \: \bar{\phi}_{RIS}^2 \sigma_0}{r_1^4 r_2^2 \: 2^{12}\: \pi^4 }|{\bf 1}_N^T \: \bSigma \: {\bf 1}_M|^4 
		\end{aligned}
	\end{eqnarray}
	and the resulting SCR can be cast as
	\begin{equation}
		{\mbox{SCR}}_{bwl}  =  \frac{4}{\pi \: r_2^2 \bar{\phi}_{RIS}^2}    \frac{\sigma}{\sigma_0} .
	\end{equation}
	The same considerations {in the pulse length case} apply except the dependence over $r_2$ {and $\phi_{RIS}$} which follow {now} a $1/r_2^2$ and $1/\phi^2_{RIS}$ law, respectively.

	\subsection{Volume clutter}
	In this sub-section, the SCR expression for volume clutter is obtained.
	Specifically, the scatterers are assumed uniformly distributed across the illuminated volume
	\begin{equation}
		\begin{aligned}
			V_c  & = \pi \: r_2^2 \left(\frac{c \tau}{2}\right) \tan{\left(\frac{\bar{\phi}_{RIS}}{2}\right)} \tan{\left(\frac{\bar{\theta}_{RIS}}{2}\right)} \\ & \simeq \frac{\pi}{4} \left(\frac{c \tau}{2}\right) r_2^2 \bar{\phi}_{RIS} \bar{\theta}_{RIS},
		\end{aligned}
	\end{equation}
	where $\bar{\theta}_{RIS}$ denotes the elevation beamwidth of the RIS while  the last approximation relies on the small angle assumption for $\bar{\phi}_{RIS}$ and $\bar{\theta}_{RIS}$. Therefore, the clutter power return is given by
	\begin{eqnarray}
		\begin{aligned}
			&C \; = \; {\frac{G_T^2 P_T G^2 {F^{tot}}^2 d_x^2 d_y^2 \eta_{RIS}^2\lambda_0^2 V_c \gamma_0}{r_1^4 r_2^4  (4\pi)^5} }|{\bf 1}_N^T \: \bSigma \: {\bf 1}_M|^4 =  \\
			& {\frac{G_T^2 P_T G^2 {F^{tot}}^2 d_x^2 d_y^2 \eta_{RIS}^2\lambda_0^2 \: c \tau \:  \bar{\phi}_{RIS} \bar{\theta}_{RIS} \: \gamma_0}{r_1^4 r_2^2  2^{13} \pi^4} }|{\bf 1}_N^T \: \bSigma \: {\bf 1}_M|^4 , \nonumber
		\end{aligned}
	\end{eqnarray}
	where $\gamma_0$ is the volume reflectivity. 
	As a consequence, the SCR for volume clutter is given by
	\begin{equation}
		{\mbox{SCR}}_v = \frac{8}{\pi \: c \: \tau \: r_2^2\:  \bar{\phi}_{RIS} \bar{\theta}_{RIS}}  \frac{\sigma}{\gamma_0},
	\end{equation}
	resulting in an inverse square law dependence on the $r_2$ distance between the RIS and the target.
	Similar to the surface clutter case, the distance between the radar and the RIS as well as the radar {transmit/receive} beamwidths do not affect the SCR.
	Finally, an inverse linear dependence with respect to both $\bar{\phi}_{RIS}$ and $\bar{\theta}_{RIS}$ is present, {since the aforementioned parameters} determine the size of the volume clutter cell.

	\section{Operation in Sub-Region 2 and Radar Echo Model}
\begin{figure*}[h!] 
		\centering
		\includegraphics[trim=0 205 5 55,clip,width=0.7\linewidth]{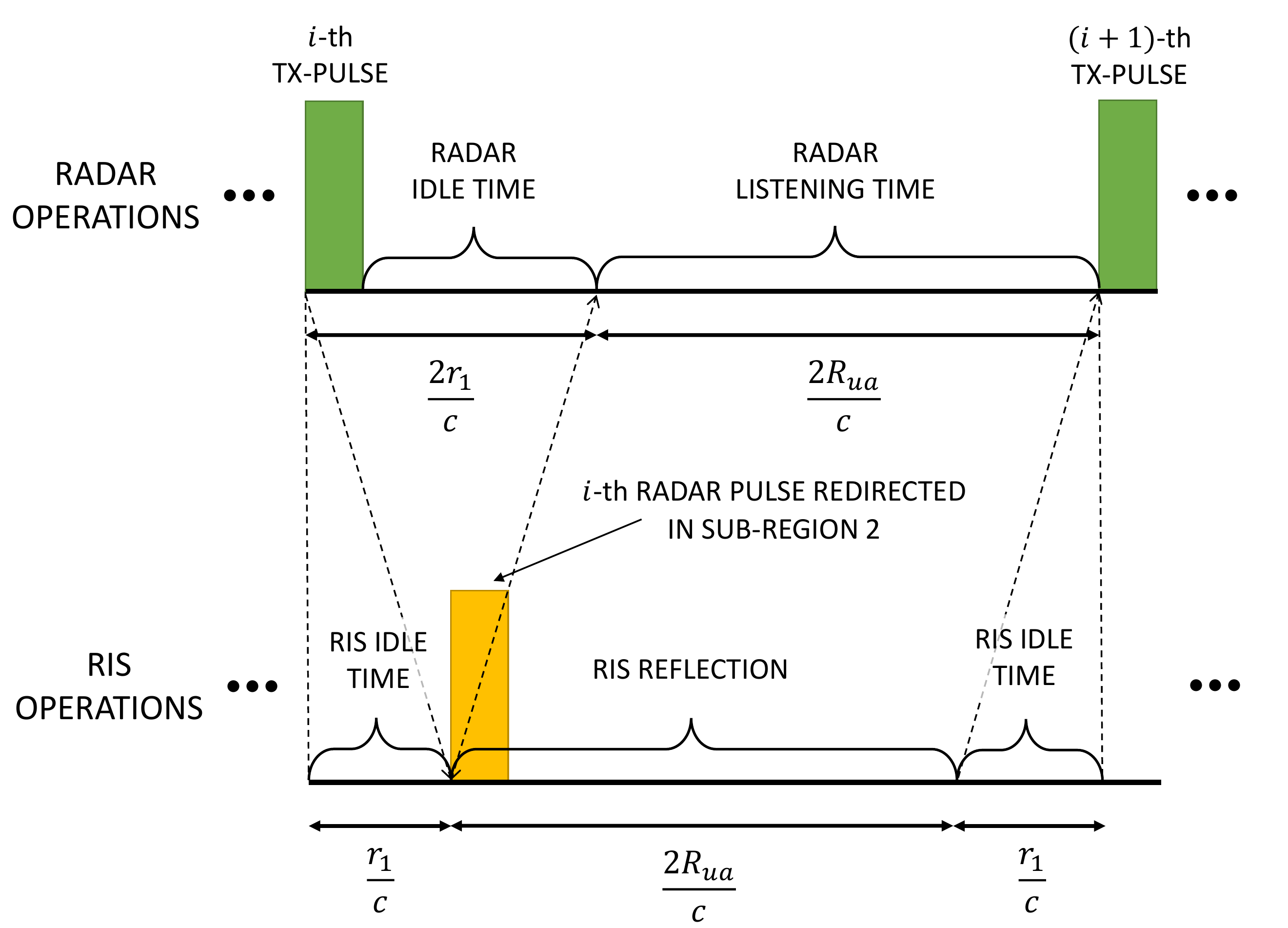}
		\caption{An illustrative example of {the} radar burst transmission for the operative mode in sub-region 2.}
		\label{fig:5} 
	\end{figure*}

	An example of a radar burst transmission for the operative mode in sub-region 2 is provided in Fig.~\ref{fig:5}. 
	The first pulse (whose duration, as already said, is denoted by $\tau$) is transmitted by the radar and reaches the RIS after a time lapse $r_1/c$. The interval $[\tau, 2\:r_1/c]$ represents an ``idle'' interval where radar transmission and reception from sub-region 2 is inhibited. The programmed surface re-directs the pulse to cover with a specific beam the desired portion of sub-region 2, via an appropriate {phase steering}. Then, a listening time starts where return signals are forwarded by the RIS toward the radar which performs data acquisition to form the fast-time slow-time matrix. The length of the listening time is {$2R_{ua}/c$ with} $R_{ua}$ the maximum unambiguous range with respect to the RIS. Then after {$2R_{ua}/c + 2 r_1/c$} the second pulse is transmitted by the radar and the cycle continues until the end of the dwell.  As discussed in {\bf Remark 1} of Section \ref{radar_eq_2_sec}, the azimuth resolution is determined by the size and geometrical{/physical} properties of the RIS as well as by the radar wavelength.
	
	{To gather insights on} the range and Doppler resolutions, the signal model {for the} echo induced by a target is now developed. To this end, let us denote by $s(t)$ the complex envelope of the narrow-band radar transmit signal, i.e.,
	$$s(t)=\displaystyle{\sum_{i=1}^{N_p}} \sqrt{P_T} p(t-(i-1)T)e^{j 2 \pi f_0 t}$$ 
	where $p(t)$ is a unit-energy baseband pulse of duration $\tau$ and bandwidth $B$, and $f_0=\frac{c}{\lambda_0}$ is the carrier frequency (with $c$ the speed of light).
	
	A prospective target at location $\bs_T$ (in sub-region 2) and velocity $\bv_T$ produces a backscattered electric field toward the RIS, given by
	\begin{eqnarray}\label{RIS_rx}
		\bE_T^{RIS}(t)=\displaystyle{\sum_{i=1}^{N_p}} \alpha_1 p(t-\tau_1-(i-1)T)e^{j 2 \pi (f_0+{f_d}_T) t} \bS_2 
	\end{eqnarray} 
	where $\tau_1=\frac{r_1+2r_2}{c}$, ${f_d}_T=2\frac{ v_{r, RIS}}{\lambda_0}$, 
	with $v_{r, RIS}=\frac{(\bs_{RIS}-\bs_T)^T }{\|\bs_{RIS}-\bs_T\|}\bv_T$  the component of the target velocity along the radial line from the RIS to the target; furthermore, $\alpha_1=\sqrt{\frac{2 Z_0{\cal{P}}\sigma}{(4\pi L_1)}} e^{j\varphi_1}$,  with  $L_1$ the loss factor involved in the considered partial propagation path and $\varphi_1$ related to the phase responses of the radar transmit antenna, the RIS-patches, the target as well as the propagation path delays. Otherwise stated, $\bE_T^{RIS}(t)$ is proportional to a delayed and frequency shifted version  of the radar transmitted burst. The electric field {in eq.}~(\ref{RIS_rx}) is thus focused toward the radar via the programmable reflecting coefficients $\bGamma$, resulting, under {the} ``stop-and-hop'' assumption, into the target-induced radar received signal
	\begin{eqnarray}\label{radar_rx}
		s_R(t)=\displaystyle{\sum_{i=1}^{N_p}} \alpha p(t-\tau_0-(i-1)T)e^{j 2 \pi (f_0+{f_d}_T) t} 
	\end{eqnarray} 
	where $\tau_0=2\frac{r_1+r_2}{c}$ and $\alpha=A e^{j\varphi}$ is a complex value with {$A=\sqrt{P^{rx}}/\sqrt{L_t L_{atm}L_r L_{ris}}$} and $\varphi$ {encompassing} all the phase terms.
	
	According to the conventional {and consolidated} radar signal processing, {the signal in eq. (\ref{radar_rx})} is down converted to baseband and undergoes a matched filtering process w.r.t. the pulse $p(t)$, i.e., pulse compression. The {pulse-compressed} output-signal is thus given by 
	\begin{equation}\label{radar_rx2}
		z_R(t) \:=\: \displaystyle{\sum_{i=1}^{N_p}} \alpha \chi_p(t-\tau_0-(i-1)T,{f_d}_T) e^{j 2 \pi {\nu_d}_T (i-1)}
	\end{equation} 
	where
	\begin{itemize}
		\item $\chi_p(t_1,f_d)=\displaystyle{\int_{-\infty}^\infty} p(t)p^*(t-t_1)e^{j 2 \pi f_d t} dt$
		is the (complex) ambiguity function of the
		pulse waveform $p(t)$,
		\item ${\nu_d}_T={f_d}_T T$ is the  normalized target Doppler frequency.
	\end{itemize}
	Expression (\ref{radar_rx2}) can be further manipulated  if $p(t)$ is Doppler tolerant, i.e., $\chi_p(t_1,{f_d}_T)\simeq \chi_p(t_1,0)$, at the Doppler {frequencies} of interest. {In this case}
	\begin{eqnarray}\label{radar_rx_1}
		z_R(t)\simeq  \displaystyle{\sum_{i=1}^{N_p}} \alpha r_p(t-\tau_0-(i-1)T) e^{j 2 \pi {\nu_d}_T (i-1)}
	\end{eqnarray} 
	where $r_p(t_1)=\chi_p(t_1,0)$ is the autocorrelation function of $p(t)$.
	
	Now, observing that the functional forms of (\ref{radar_rx2}) and (\ref{radar_rx_1}) are the same as those {for} a standard monostatic radar, it follows that two targets in region 2, located at ${{\bs}_{T}}_1$ and ${{\bs}_{T}}_2$ with Doppler frequencies ${{f_d}_{T}}_1$ and ${{f_d}_{T}}_2$, {respectively}, can be distinguished in range if 
	$$\big|(\| {{\bs_{T}}_1} - \bs_{ris} \| + r_1 )  - (\| {{\bs_{T}}_2}- \bs_{ris}\| + r_1 )\big|=|r_2^1-r_2^2|\geq \frac{c}{2B},$$
	and in Doppler if
	$$|{f_d}_{T, 1}-{f_d}_{T,2}|\geq \frac{1}{N_p T},$$
	with $r_2^k$ the distance between the RIS and the $k$-th target, $k=1,2$
	Otherwise stated, the range and Doppler resolutions of the RIS-assisted radar are $\Delta R_{ris}= \frac{c}{2B}$ and $\Delta {f_d}_{ris}=\frac{1}{N_p T}$, respectively, {where the target range $r_1+r_2$ is composed by the fixed and known bias term $r_1$ plus the range $r_2=\|\bs_T-\bs_{ris} \|$ computed w.r.t.  $\mbox{CRS}_2$,  and Doppler frequency refers to the target radial velocity w.r.t.  $\mbox{CRS}_2$}. As a consequence, as in sub-region 1,
	the range resolution in sub-region 2 is ruled by the bandwidth $B$ of the radar waveform. Moreover, the Doppler resolution is inversely proportional to the dwell time $N_p T$.
	
	Before concluding this section, let us observe that eq. (\ref{radar_rx_1}), together with the derived range resolution, allow {for} the construction of the fast-time slow-time data matrix $\bD$ corresponding to sub-region 2. Specifically, let $\bar{z}_R(t)$ be the radar received signal after down-conversion and pulse compression. Then the $(h,l)$-th entry of $\bD$, $h=1,\ldots,h_{max}+1$, $l=1,\ldots N_p$, is given by
	$$\bD(h,l)=\bar{z}_R\left(2 \frac{r_1+r_2^b}{c}+\frac{(h-1)}{B}+(l-1)T\right),$$
	where $r_2^b$ is the {minimum operative range} in sub-region 2 and $h_{max}$ such that $r_1+r_2^b+  h_{max} \Delta R_{ris}=R_{ua}$.
	
	This represents the data block used by the {radar} to perform any successive signal processing such as MTI, pulse Doppler, CFAR, etc.

	\section{Numerical Results}
	In this section, numerical examples are provided to assess the performance of the proposed N-LOS modality in terms of SNR and $\text{P}_\text{d}$. Besides, an analysis on the SNR loss {w.r.t. a clairvoyant monostatic configuration} is provided {too}.
	%le altre condizioni
	To this end, three different RISs (each comprising $N\times M$ elements) with {$N=M \in \{101,133,201\}$}, are considered. Therefore, assuming an inter-element space of {$\lambda_0/2 = 0.015~\text{m}$}, {the} area {occupied by each} of the considered RISs {is} {approximately} equal to $1.5\:\text{m} \times 1.5\:\text{m}$, $2.0\:\text{m} \times 2.0\:\text{m}$, and $3.0\:\text{m} \times 3.0\:\text{m}$, respectively. As a consequence, the corresponding far field distance (FFD) is {$152.91$ m, $265.15$ m, and $605.60$ m, respectively,} where FFD is {computed as} $r_{FFD} = 2 (\max(d_x\,N, \; d_y\,M))^2/\lambda_0$. The normalized RIS-patch power radiation pattern {is modeled as}
	\begin{equation*}
		F(\theta_R)=	\begin{cases}
			\cos^{1.5}(\theta_R) & \text{ if } \theta_R \in [0, \pi/2] \\ 
			0 & \text{ otherwise }
		\end{cases} ,
	\end{equation*}
	which is suitable for an {antenna isotropic in azimuth,} operating in an {elevation} range (w.r.t. the  RIS) of $[0, \pi/3]$~\cite{richards1}.
	As to the target, a typical micro UAV, characterized by a RCS of $0.02~\text{m}^2$, is considered.
	The values of the {system} parameters involved in the analyzed case studies {are summarized} in Table~\ref{tab-1}.
	\begin{table}[h!]
		\centering
		\caption{Simulation Parameters}\label{tab-1}
		\begin{tabular}{@{}cc@{}}
			\hline
			\hline
			\textbf{Parameter} & \textbf{Value}\\
			\hline
			$N,M$            & $\{101, 133, 201\}$                 \\
			$d_x, d_y$       & $\lambda_0 / 2$                   \\
			$\eta_{RIS}$     & $0.8$                              \\
			$\sigma$         & $0.02~\text{m}^2$ \\
			$r_1$            & $1000$ m                          \\
			$f$              & $10$ GHz                          \\
			$N_p$            & $\{8,16,32, 64, 128\}$                       \\
			$\tau$           & $1.5 \times 10^{-6}$ s                   \\
			$P_t$            & $26$ dBW                           \\
			$G_t$            & $38$ dB                           \\
			$G$              & $4$ dB                            \\
			$L_{tot}$        & $6$ dB                            \\	
			$F_N$              & $2.5$ dB                          \\
			$F^R(\theta_R,\phi_R)$        			& $0$ dB                            \\
			\hline
		\end{tabular}
	\end{table}
	Moreover, Table~\ref{tab-2} reports the dwell times (assuming a maximum unambiguous range of $10$ Km, {i.e., a Pulse Repetition Time (PRT) of $66.71$ $\mu$s}) corresponding to the different number of pulses used to evaluate the SNR and $\text{P}_\text{d}$. {All of them are compatible with a range resolution of $15$ m which can be obtained with a modulated pulse whose bandwidth $B$ is $10$ MHz.} {It is also important to remark that $B$ poses a constraint to the size of the RIS because of the narrowband assumption which can be no longer met when $\max(d_x\, N,d_y\, M)>3.0$ m.} 
	
	\begin{table}[h!]
		\centering
		\caption{Number of Pulses and dwell time assuming an unambiguous range of $10$Km}\label{tab-2}
		\begin{tabular}{@{}cc@{}}
			\hline
			\hline
			\textbf{Number of Pulses} & \textbf{Dwell Time}\\
			\hline
			$8$       	& $0.53$ ms  		\\
			$16$       	& $1.07$ ms		\\
			$32$     	& $2.13$ ms 		\\
			$64$     	& $4.27$ ms		\\
			$128$     	& $8.54$ ms     \\
			\hline
		\end{tabular}
	\end{table}
	
	% --------------------------------------
	
	\begin{figure*}[htbp] 
		\centering
		\subfloat[\label{fig:patterns_1}]{%
			\includegraphics[trim=130 45 50 20,clip,width=0.32\linewidth]{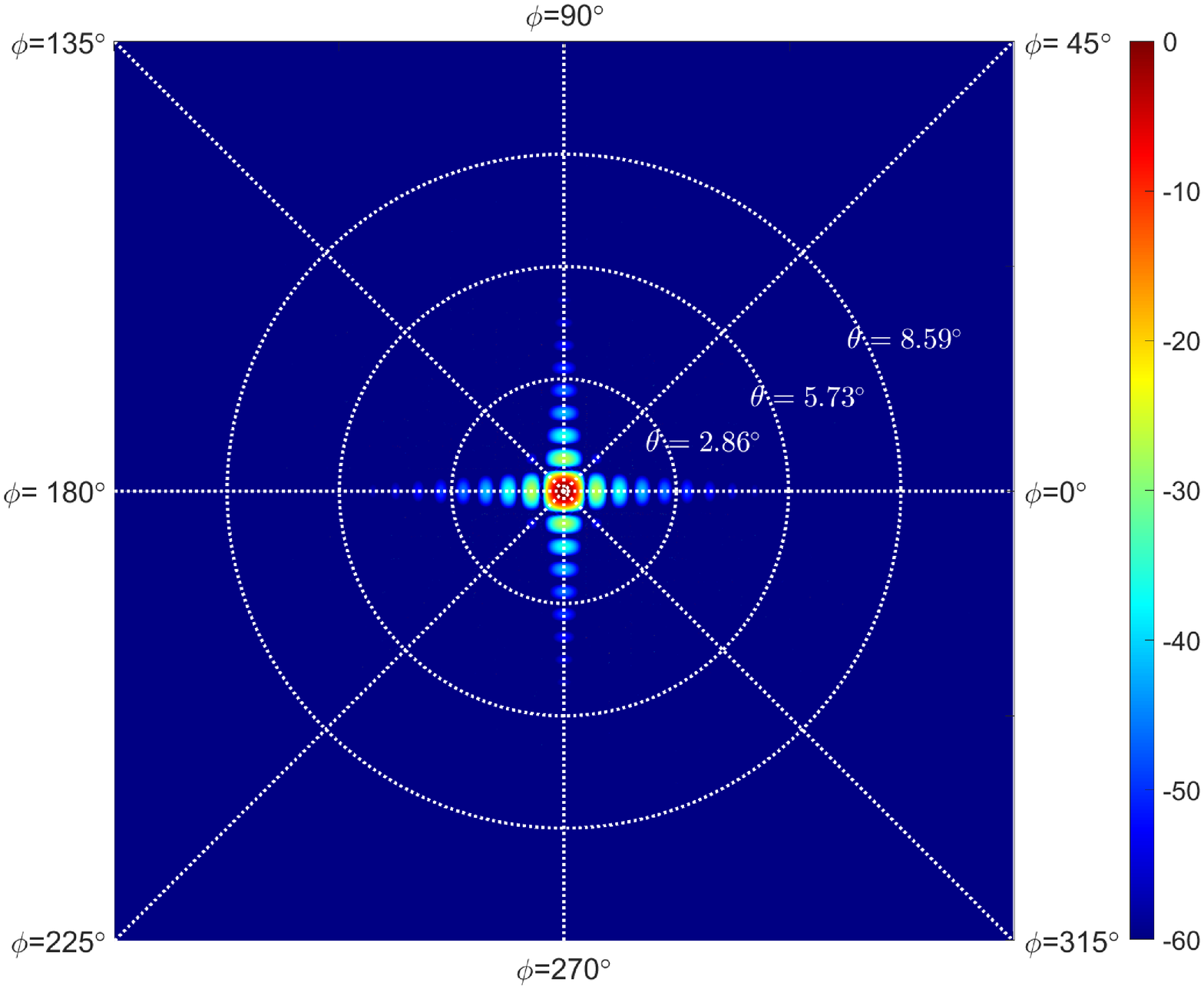}}
		\hspace{10pt}\subfloat[\label{fig:patterns_2}]{%
			\includegraphics[trim=130 45 50 20,clip,width=0.32\linewidth]{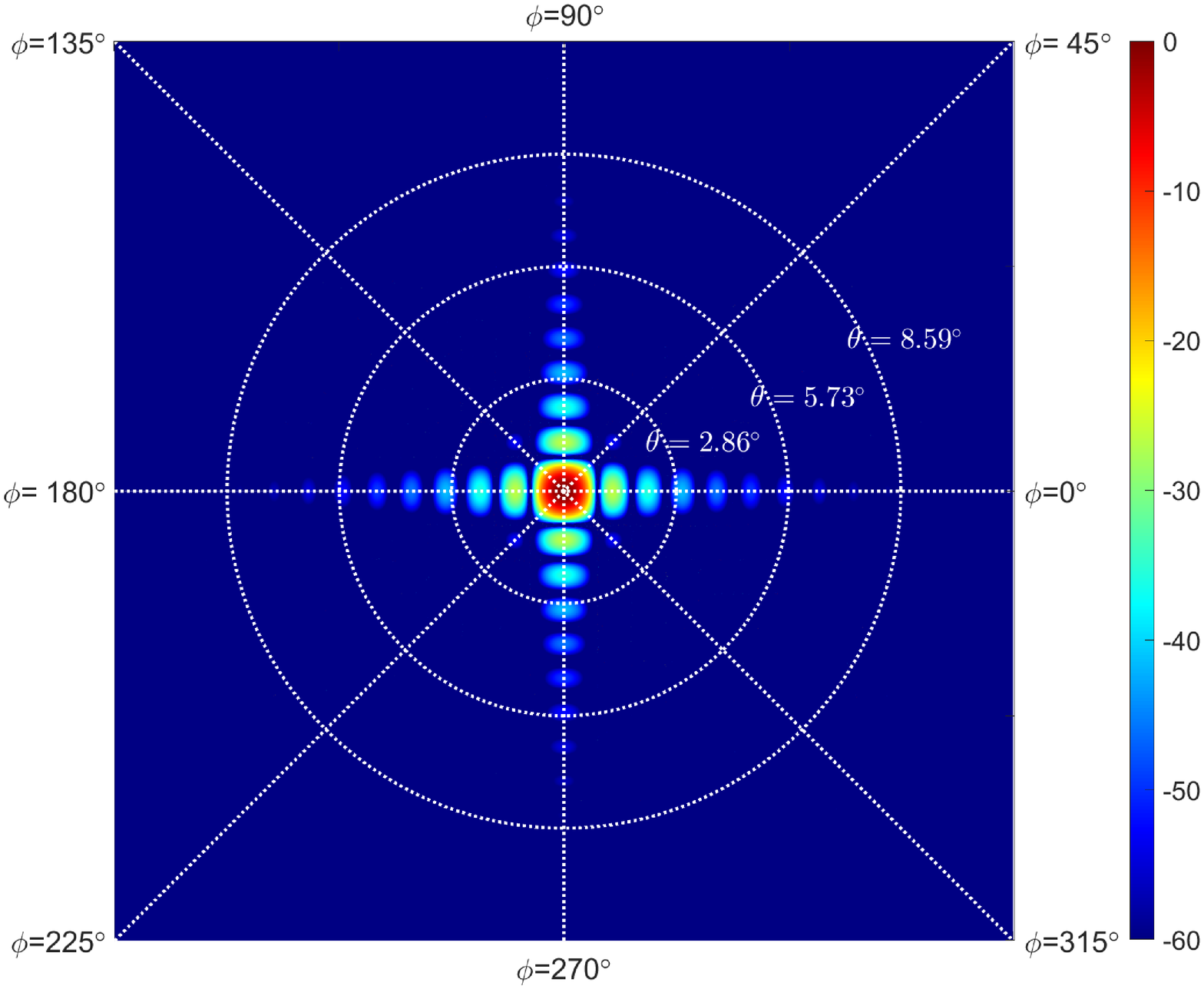}}
		\hspace{10pt}\subfloat[\label{fig:patterns_3}]{%
			\includegraphics[trim=130 45 50 20,clip,width=0.32\linewidth]{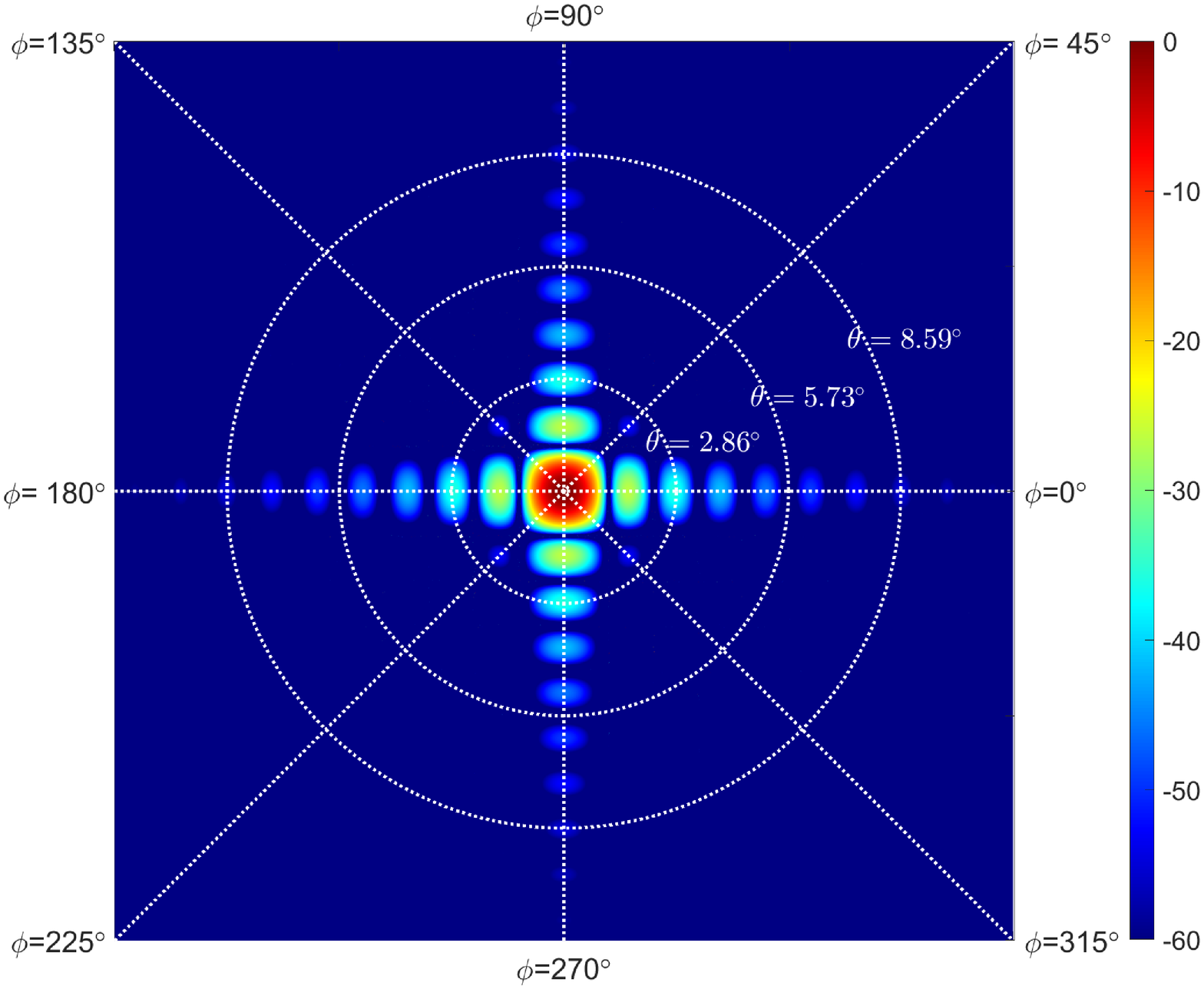}}
		\\
		\subfloat[\label{fig:patterns_4}]{%
			\includegraphics[trim=130 45 50 20,clip,width=0.32\linewidth]{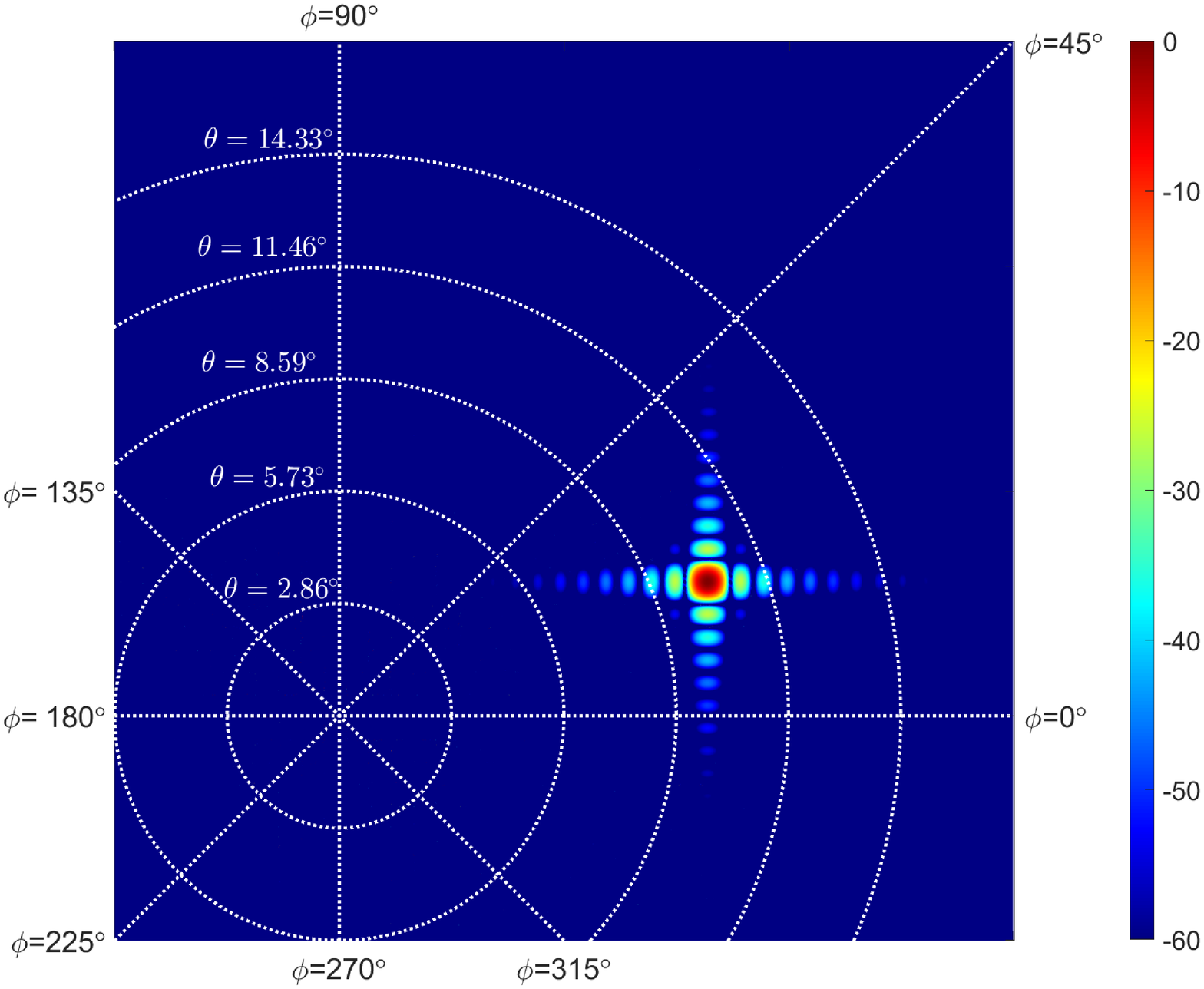}}
		\hspace{10pt}\subfloat[\label{fig:patterns_5}]{%
			\includegraphics[trim=130 45 50 20,clip,width=0.32\linewidth]{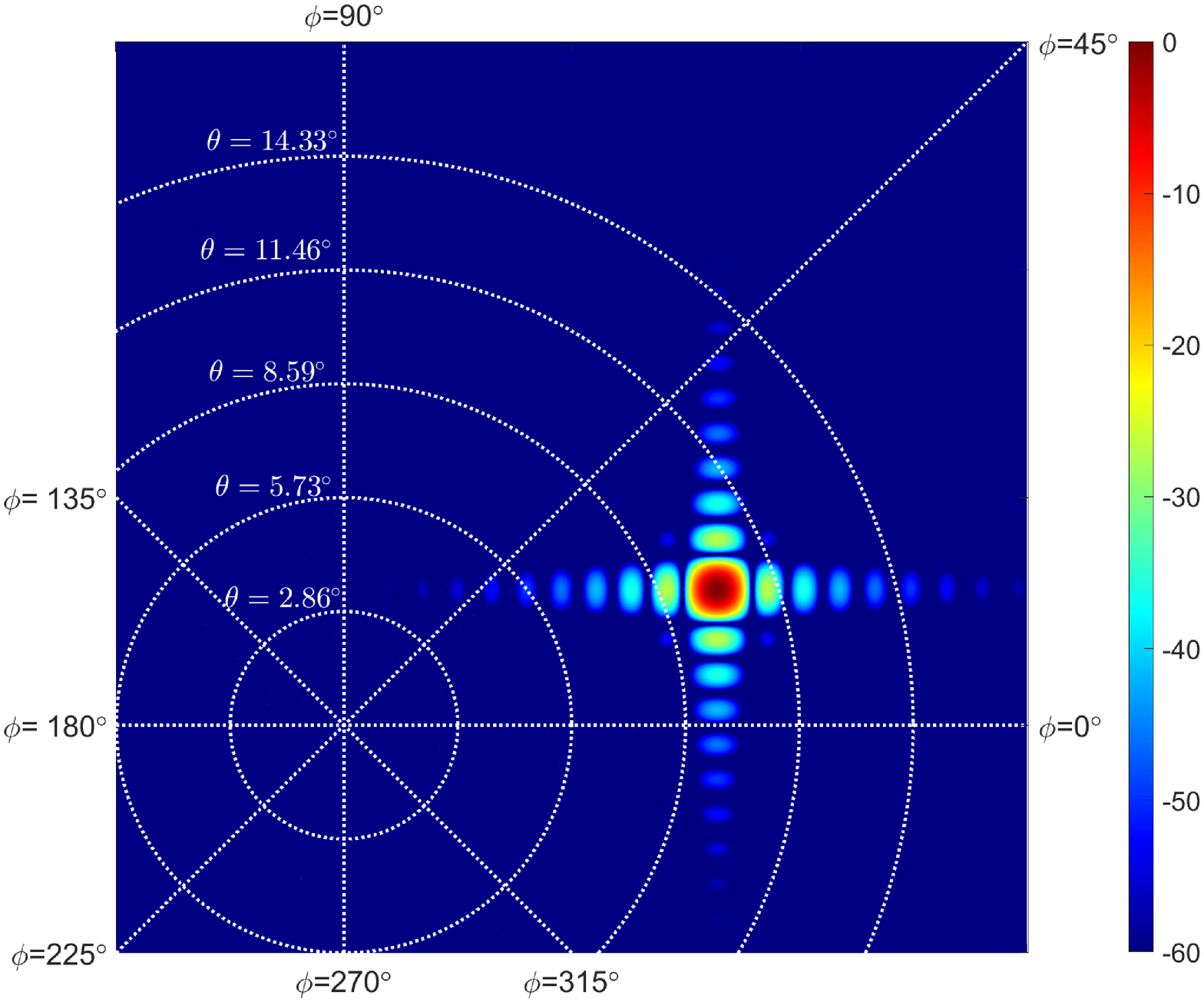}}
		\hspace{10pt}\subfloat[\label{fig:patterns_6}]{%
			\includegraphics[trim=130 45 50 20,clip,width=0.32\linewidth]{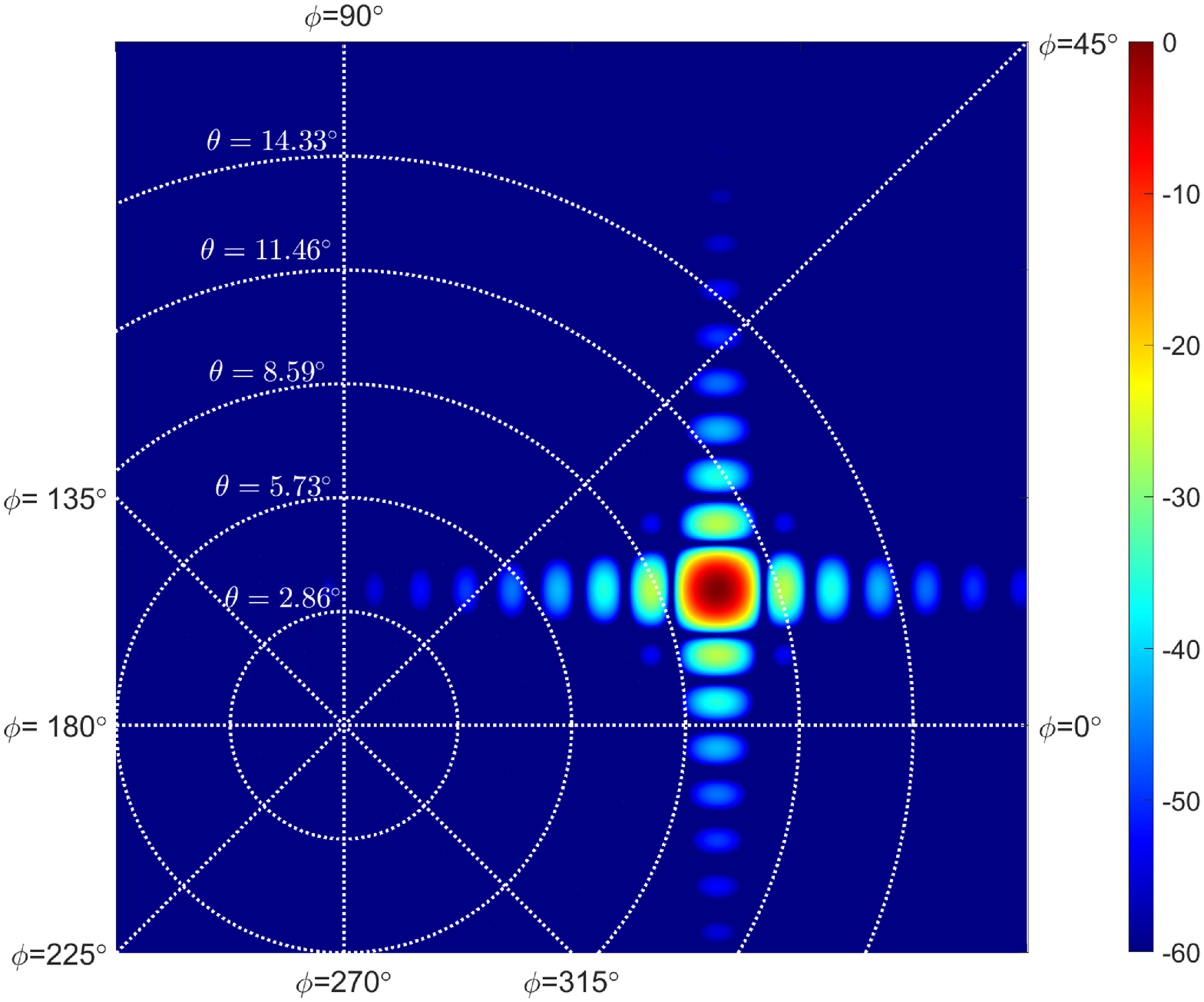}}
		\caption{Normalized RIS induced pattern versus $\theta_{RIS}^{T_a}$ and $\phi_{RIS}^{T_a}$ assuming $r_1 = 1000$ m and $r_2 = 1500$ m. 
			Figs.~\subref{fig:patterns_1} and \subref{fig:patterns_4} refer to $N=M=201$,  Figs.~\subref{fig:patterns_2} and \subref{fig:patterns_5} $N=M=133$, whereas  Figs.~\subref{fig:patterns_3} and \subref{fig:patterns_6} $N=M=101$.
			Besides, $\theta_2 = \phi_2 = 0^\circ$ is considered in Figs.~\subref{fig:patterns_1}, \subref{fig:patterns_2} and \subref{fig:patterns_3}, whereas  $\theta_2 = 10^\circ$, $\phi_2 = 20^\circ$ is assumed in Figs.~\subref{fig:patterns_4}, \subref{fig:patterns_5} and \subref{fig:patterns_6}.}
		\centering
		\label{fig:patterns} 
	\end{figure*}
	Fig.~\ref{fig:patterns} reports the RIS induced pattern {pertaining to}~\eqref{patter_exp} (normalized to {the peak value} $(NM)^4$) versus $\theta_{RIS}^{T_a}$ and $\phi_{RIS}^{T_a}$, for three {different} RIS sizes (see Table~\ref{tab-1}) and two {diverse} pointing directions in sub-region 2. {Specifically, in all the plots the phase-matching condition holds true in sub-region 1, i.e., $\theta_1 = \theta_{RIS}^R$ and $\phi_1 = \phi_{RIS}^R$, and Figs.~\subref*{fig:patterns_1}, \subref*{fig:patterns_2}, and \subref*{fig:patterns_3} refer to $\theta_2 = \phi_2 = 0$, whereas Figs.~\subref*{fig:patterns_4}, \subref*{fig:patterns_5}, and \subref*{fig:patterns_6} correspond to $\theta_2 = 10^\circ$, $\phi_2 = 20^\circ$.}
	Furthermore, Figs.~\subref*{fig:patterns_1} and \subref*{fig:patterns_4} consider $N=M=201$, {while} $N=M=133$ and $N=M=101$ are {exploited} in Figs.~\subref*{fig:patterns_2},\subref*{fig:patterns_5} and Figs.~\subref*{fig:patterns_3},\subref*{fig:patterns_6}, respectively.

	Inspection of the figures {unveils} that the maximum {pattern} value is achieved when the target is located along the RIS pointing direction, in accordance {with} the theoretical {achievements} of Section IV.
	{It can {also} be observed that, regardless of the RIS size,  the beamwidths widen as the RIS is steered toward a direction far from the boresight, a well-known phenomenon in Active Electronically Scanned Arrays (AESAs)~\cite{farina1992antenna, vantrees4}. {Lastly,} the larger the RIS size the narrower the beam in sub-region 2, being the reflected energy focused better and better in the desired looking direction. }

	\begin{figure}[htbp] 
		\centering
		\includegraphics[width=0.98\linewidth]{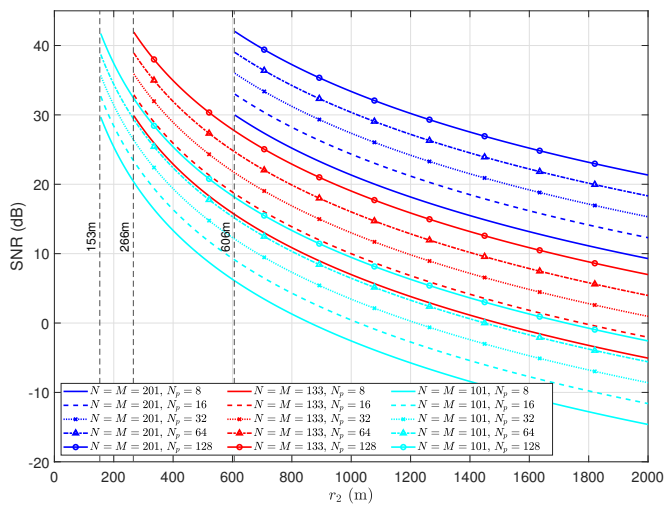}
		\caption{SNR versus $r_2$ {for different RIS sizes, number of pulses, and the parameters of Table~\ref{tab-1}}. {Moreover, $\theta_1 = \theta_{RIS}^R=30^\circ$, $\phi_1 = \phi_{RIS}^R=20^\circ$, $\theta_2 = \theta_{RIS}^{T_a}=0^\circ$, and $\phi_1 = \phi_{RIS}^{T_a}=0^\circ$.}}
		\label{fig:snr_vs_r2} 
	\end{figure}
	\begin{figure*}[htbp] 
		\centering
		\includegraphics[trim=90 20 80 20,clip,width=0.85\linewidth]{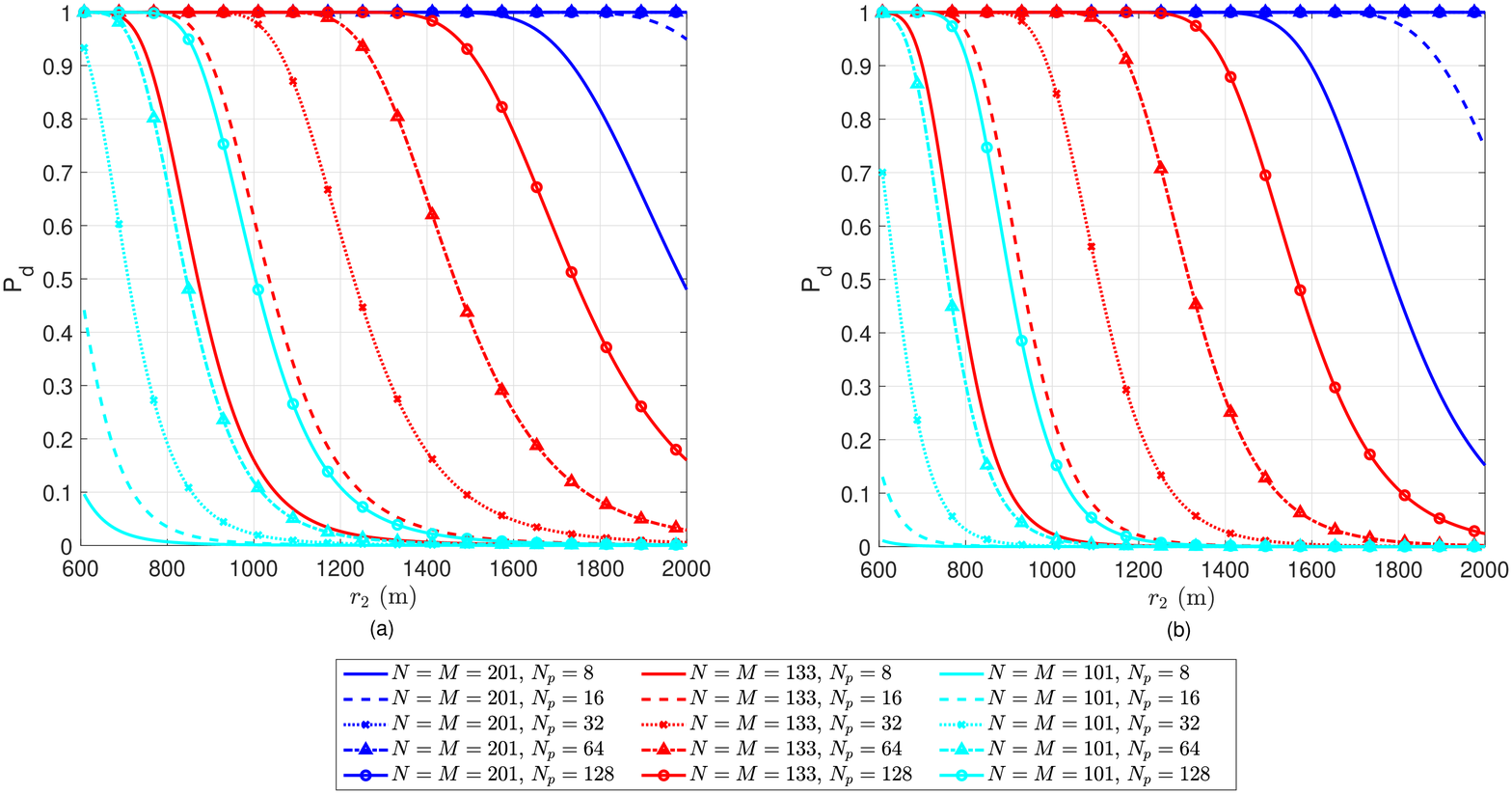}
		\caption{$\text{P}_\text{d}$ versus $r_2$ for a SW0 target case {and different RIS sizes as well as number of pulses (see Table~\ref{tab-1})}. {Therein, $\theta_1 = \theta_{RIS}^R=30^\circ$, $\phi_1 = \phi_{RIS}^R=20^\circ$, $\theta_2 = \theta_{RIS}^{T_a}=0^\circ$, $\phi_1 = \phi_{RIS}^{T_a}=0^\circ$, and} in (a) $\text{P}_{FA} = 10^{-4}$ while in (b) $\text{P}_{FA} = 10^{-6}$. }
		\label{fig:pd_sw0}  
	\end{figure*}
	
	\begin{figure*}[htbp] 
		\centering
		\includegraphics[trim=90 20 80 20,clip,width=0.85\linewidth]{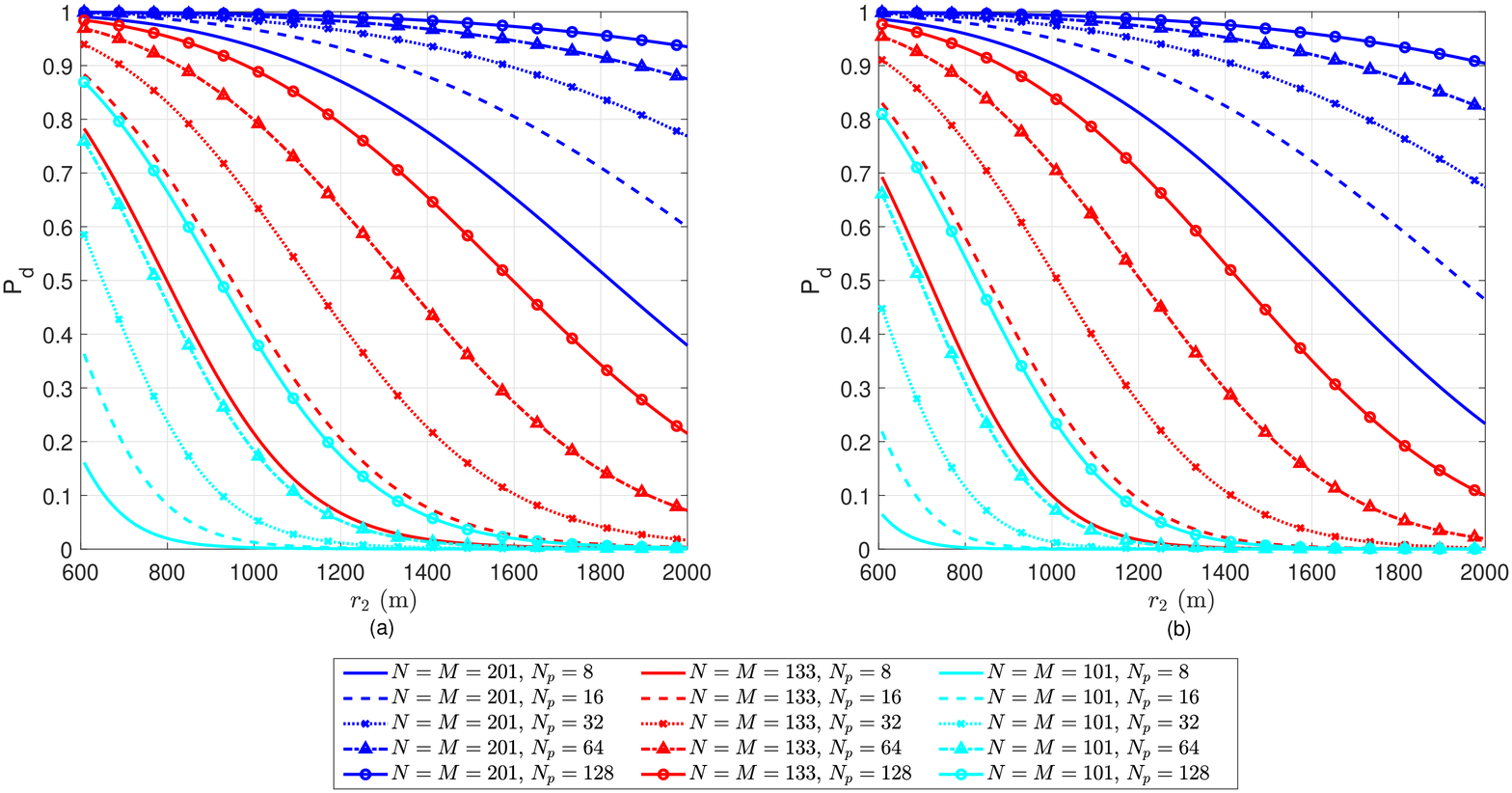}
		\caption{$\text{P}_\text{d}$ versus $r_2$ for a SW1 target case {and different RIS sizes as well as number of pulses (see Table~\ref{tab-1})}. {Therein, $\theta_1 = \theta_{RIS}^R=30^\circ$, $\phi_1 = \phi_{RIS}^R=20^\circ$, $\theta_2 = \theta_{RIS}^{T_a}=0^\circ$, $\phi_1 = \phi_{RIS}^{T_a}=0^\circ$, and} in (a) $\text{P}_{FA} = 10^{-4}$ while in (b) $\text{P}_{FA} = 10^{-6}$. }
		\label{fig:pd_sw1} 
	\end{figure*}
	
	Fig.~\ref{fig:snr_vs_r2} displays the ${\mbox{SNR}}_c(N_p)$ (defined in~\eqref{eq_8}) versus $r_2$ assuming {the} three different RIS {configurations} of Table~\ref{tab-1} and different choices for the number of pulses. The corresponding FFD ranges are also {superimposed to the figure}.
	Not surprisingly, the SNR decreases as the distance between the target and the RIS increases, whereas an improvement is connected {with} a larger number of {RIS} elements as well as to an increased number of {coherently integrated} pulses.
	{Nevertheless}, a large RIS area corresponds to an increased FFD as well as to a more complex RIS control. {On one hand} this underpins a trade-off between the SNR level and the {complexity of the} scenario {while on the other it} suggests the {development} of new radar processing strategies for near-field operation if large RIS dimensions are required. Furthermore, a crucial role in the SNR budget is played by the number of pulses (and hence by the dwell time on the target) with SNR improvement up to $12$ dB {when} $N_p=128$ w.r.t. the $N_p = 8$ case. {Finally, if a mini UAV ($\sigma=0.2~\text{m}^2$) is considered as prospective target in place of the micro UAV, an improvement of $10$ dB in the resulting SNR occurs.}

	Figs.~\ref{fig:pd_sw0} reports the $\text{P}_\text{d}$ versus $r_2$ for a non fluctuating target, i.e., the target RCS is a deterministic parameter (also called a Swerling 0 (SW0) {model}), whereas Fig.~\ref{fig:pd_sw1} refers to a SW1 target, i.e., the RCS is drawn from {an exponential distribution} with a scan-to-scan decorellation. Figs.~\ref{fig:pd_sw0}{(a)} and \ref{fig:pd_sw1}{(a)} refer to $\text{P}_{FA} = 10^{-4}$, whereas Figs.~\ref{fig:pd_sw0}{(b)} and \ref{fig:pd_sw1}{(b)} consider $\text{P}_{FA} = 10^{-6}$.
	Specifically, denoting by $Q_M$ the Generalized Marcum Q function, the $\text{P}_\text{d}$ for the two analyzed cases is given by
	\begin{equation}
		\text{P}_{\text{d}, SW0} = Q_M\left(\sqrt{2 \; {\mbox{SNR}}_c}, \sqrt{-2 \; \ln(\text{P}_{FA})}\right)
	\end{equation}
	and
	\begin{equation}
		\text{P}_{\text{d}, SW1} = {\text{P}_{FA}}^{(1/(1+ {\mbox{SNR}}_c ))},
	\end{equation}
	respectively.
	
	{As expected, the plots highlight that the higher $r_2$, the lower the  $\text{P}_\text{d}$ being ${\mbox{SNR}}_c$ worse and worse. }
	{Moreover, the detectability of a SW1 target is more challenging than for a SW0 target. For instance, assuming $\text{P}_{FA} = 10^{-4}$, $N=M=133$, and $N_p = 64$, $\text{P}_\text{d}$ levels greater than $0.9$ are achievable up to $1250$ m, whereas a maximum range reduction close to $500$ m is experienced for a SW1 counterpart.}
	Besides, as $\text{P}_\text{d}$ decreases, the performance gap between the SW0 and SW1 cases {reduces} {and (as already well known) for low detection rates the fluctuation is beneficial}.
	{Finally}, assuming $\text{P}_{FA} = 10^{-6}$ {determines} a further performance degradation of about $100$ m at $\text{P}_\text{d} = 0.9$ w.r.t. the $\text{P}_{FA} = 10^{-4}$ case.
	
	\begin{figure}[htbp] 
		\centering
		\subfloat[\label{fig:LSNR_vs_r2_500m}]{%
			\includegraphics[width=0.9\linewidth]{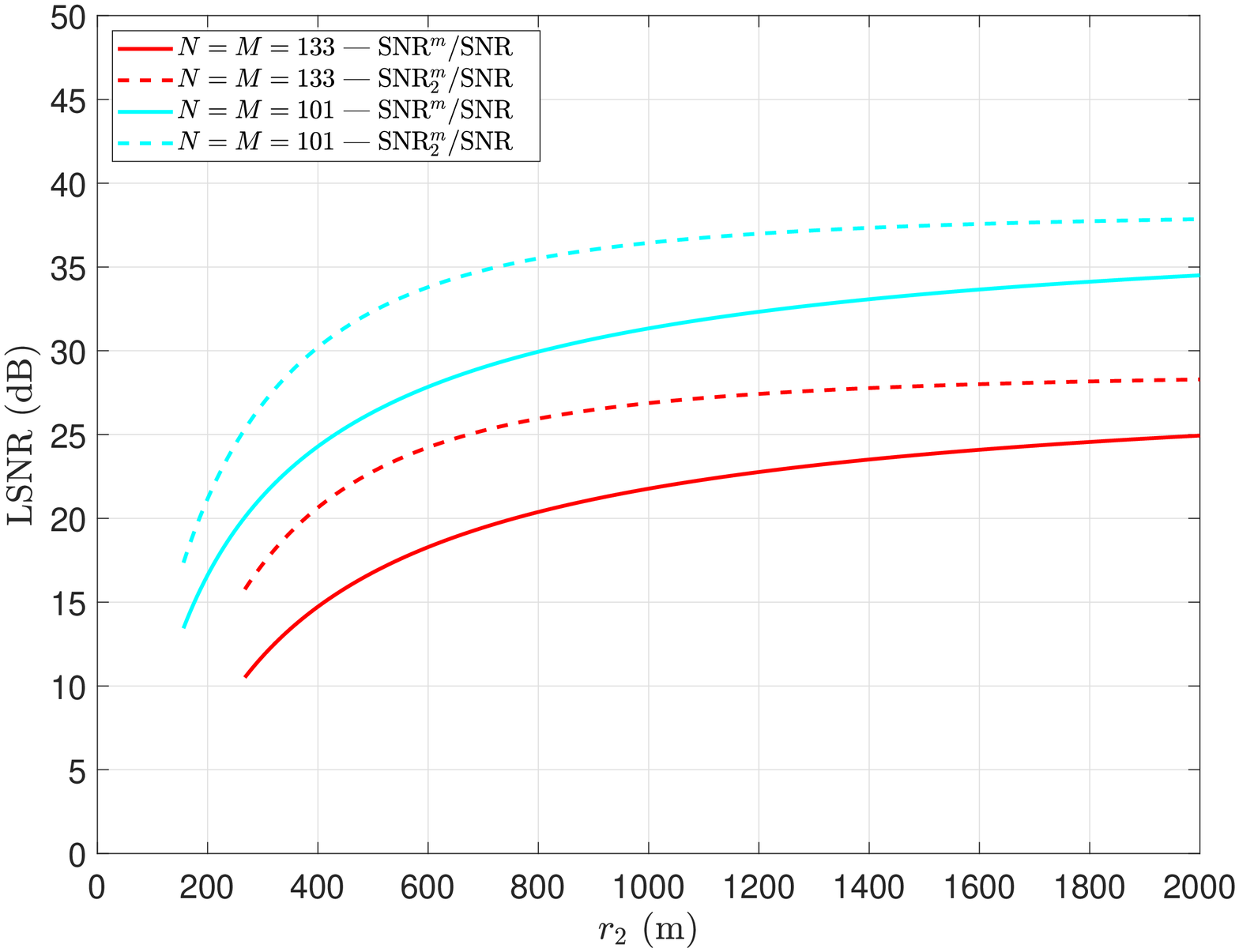}}
		\\ \subfloat[\label{fig:LSNR_vs_r2_750m}]{%
			\includegraphics[width=0.9\linewidth]{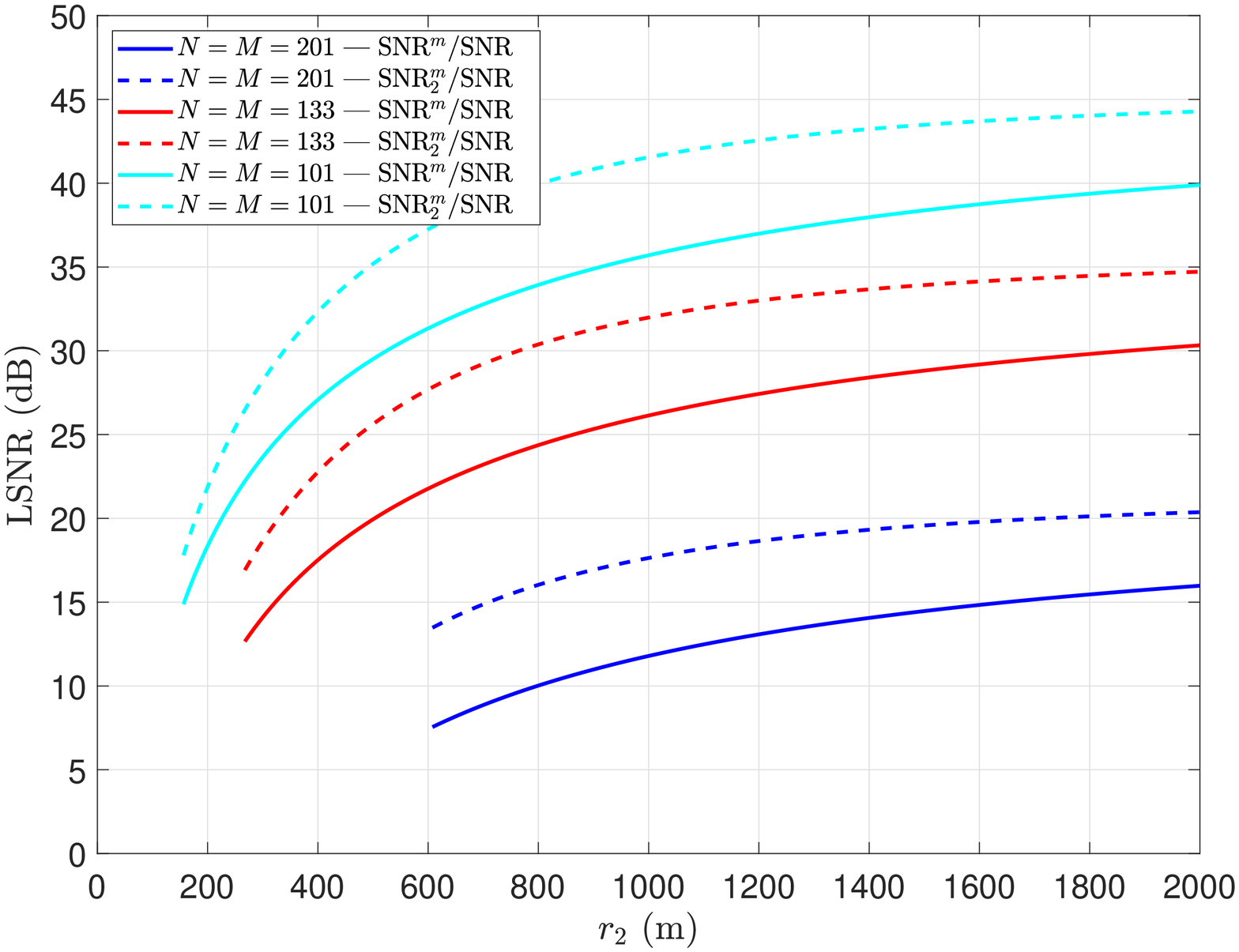}}
		\\ \subfloat[\label{fig:LSNR_vs_r2_1000m}]{%
			\includegraphics[width=0.9\linewidth]{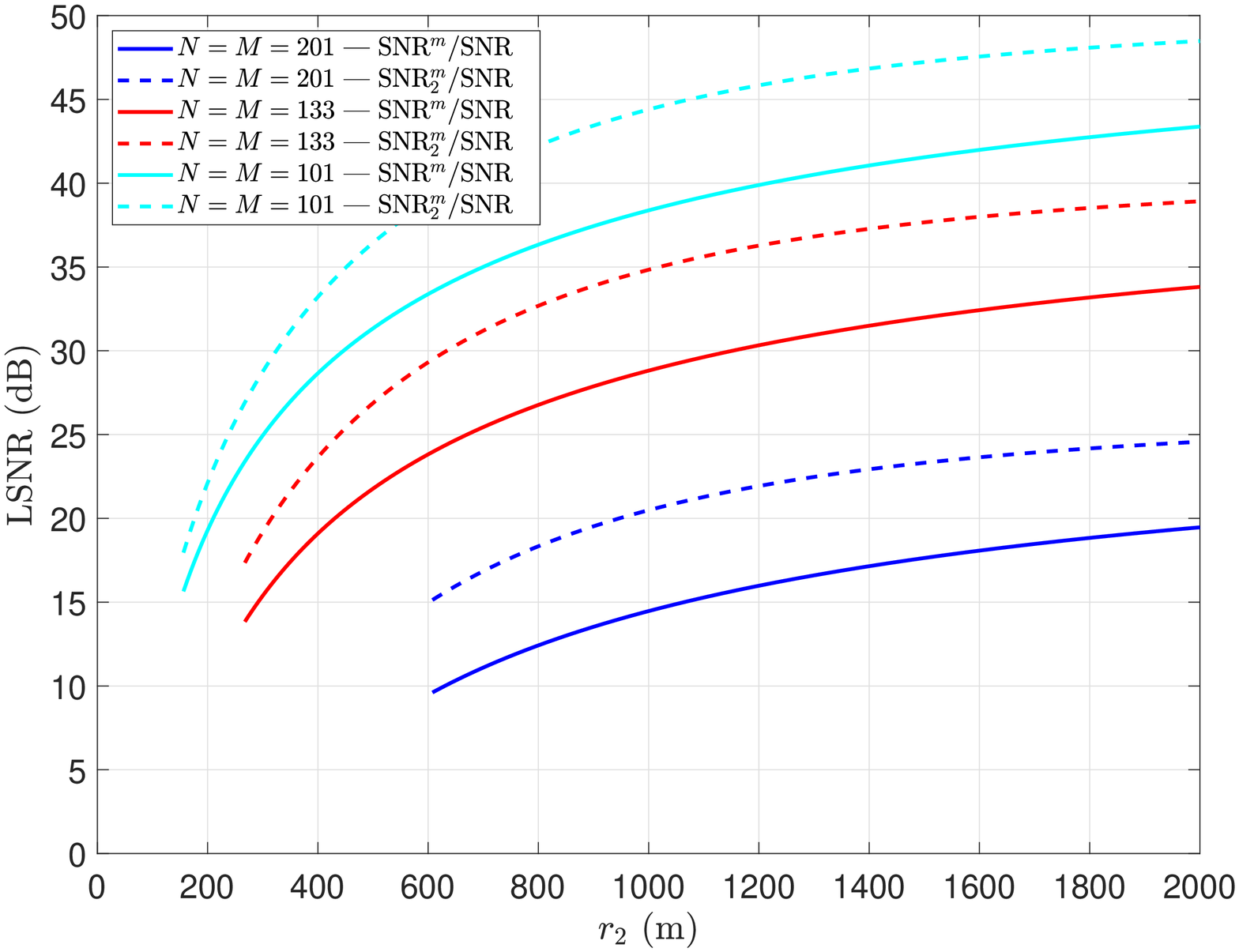}}
		\caption{LSNR versus $r_2$ assuming (a) $r_1=500$ m, (b) $r_1 = 750$ m, and (c) $r_1 = 1000$ m, different RIS sizes and parameters of Table~\ref{tab-1}. {In (a), the case $N=M=301$ is not reported because for $r_1=500$ the RIS is no longer in the far-field region of the radar.}}
		\label{fig:LSNR_vs_r2} 
	\end{figure}
	
	{Fig.~\ref{fig:LSNR_vs_r2} displays the SNR Loss ($\mbox{L}_{SNR}$) w.r.t. a clairvoyant monostatic configuration, that assumes the absence of the blockage (i.e, the presence of the LOS path). Specifically, $\mbox{L}_{SNR} = \mbox{SNR}^m/\mbox{SNR}$, with $\mbox{SNR}^m$ {pertaining to} the clairvoyant system, is computed considering two different range scenarios. The former entails a target located at a range $r_1+r_2$ and the corresponding $\mbox{SNR}^m$ is denoted {by} $\mbox{SNR}_1^m$. The latter considers a target positioned at $\sqrt{r_1^2 + r_2^2}$ and the related $\mbox{SNR}^m$ is indicated as $\mbox{SNR}_2^m$.}

	{Also}, Fig.~\subref*{fig:LSNR_vs_r2_500m}, \subref*{fig:LSNR_vs_r2_750m}, and \subref*{fig:LSNR_vs_r2_1000m} refer to $r_1 = 500, 750, 1000$ m, respectively.
	As expected, the farther the target, the larger the $\mbox{L}_{SNR}$ in all the analyzed cases. {Clearly,} regardless of the distance between the radar and the RIS, {a somewhat significant loss is experienced w.r.t. the clairvoyant monostatic system.} 
	{This is an important point and is a consequence of the two-way double-hop link established via the {RIS link and surface parameters}. {Consequently,} this {could pose} a limit to the maximum achievable radar detection range and relegates the new operational mode to {quite} short-range applications, especially for small size RISs.}
	On the other hand, inspection of the figure corroborates that the use of a large RIS reduces the $\mbox{L}_{SNR}$ with the drawback of an increased FFD distance and {a higher} RIS-controller complexity.
	
	\begin{figure}[htbp] 
		\centering
		\includegraphics[trim=45 2 60 20,clip,width=0.9\linewidth]{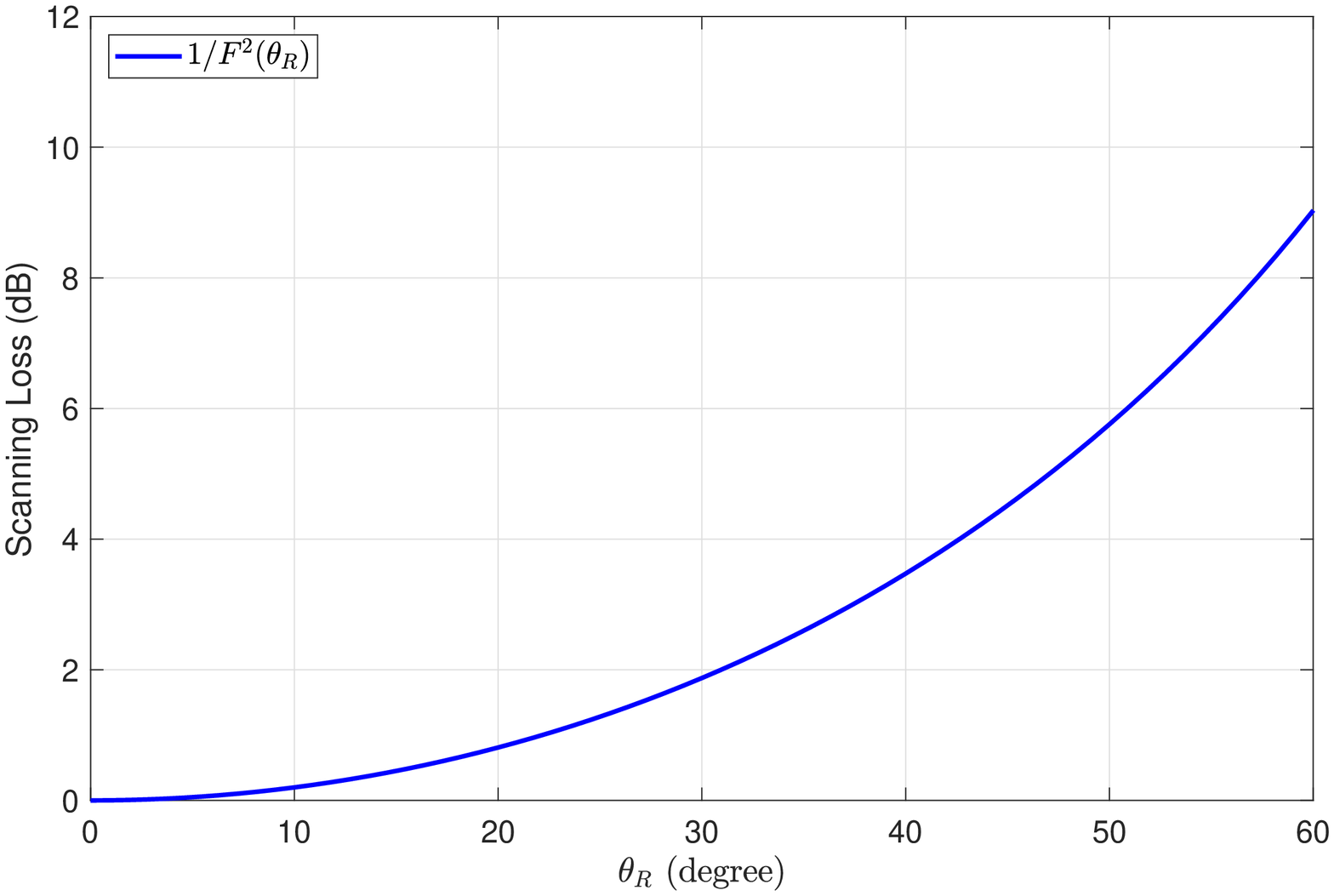}
		\caption{Scanning Loss versus $\theta_{R}$.}
		\label{fig:scanningLoss} 
	\end{figure}
	
	{The last analysis in} Fig.~\ref{fig:scanningLoss} {evaluates} the Scanning Loss $1/F^2(\theta_R)$. The {curve highlights} the capability of the RIS to scan elevation angles up to $45^\circ$ with a {loss smaller than} $5$ dB. Specifically, for $\theta_R \le 20^\circ$ the loss is {quite} negligible, otherwise a proper compensation {via} an increased dwell time {(dwell-time diversity)} {has to be granted} to guarantee the desired performance level.
	
	\section{Conclusion}\label{section:conclusion}
	In this paper the use of RIS technology is proposed to address radar surveillance in N-LOS conditions. A new sensing mode is developed via the formation of an artificial and favorable propagation environment established via the modulation of RIS parameters. For this operational regime the radar equation is laid down accounting for the artificially induced two-way and double-hop channel as well as the effects of the reflecting surface. Expressions for SNR and SCR (both for surface and volume clutter) are determined. Besides, the data acquisition procedure for N-LOS operation is discussed together with the resolution issues in
	the range, angle, and Doppler domains. A numerical analysis is carried on 
	in terms of SNR, detection performance, and SNR loss with respect to
	a LOS monostatic geometry. The impact of the RIS size and system parameters
	is assessed corroborating the theoretical capability of the new 
	framework to handle N-LOS {short-range} scenarios. 
	
	Nevertheless, it is worth observing that,
	albeit the potentialities, it is still necessary a complete understanding
	of the drawbacks connected with the employment of RIS which is
	a technology still in a maturing stage. Needless to say, even if plenty of
	research has been developed especially within the communication community,
	{further efforts are required to ponder the pros and cons.}
	Undoubtedly, it should be reckoned that from a theoretical point of view the use of RIS can represent a  fertile research field giving new degrees of freedom for system optimization. As a consequence, the RIS topic is expected ({in the} authors' opinion) to flourish also in the radar context.  In this respect, there are {diverse} radar applications which {could} benefit from this paradigm. Among them it is of interest to consider the joint use of natural multipath and the artificial (ad-hoc) multipath determined by the RIS to boost the performance of the ``around the corner radar"~\cite{8640240}. Besides, RIS can also represent key elements for the cognitive radar architecture~\cite{cognitive_book}, i.e. another level of flexibility which can be adapted based on the perception-action cycle. In other words, by leveraging perception outputs, the radar can change the RIS parameters establishing a favorable propagation{/sensing} scenario via specific target-clutter illuminations both in terms of beam steering (size and direction) and polarization of the electromagnetic wave.

	\appendix
	\subsection{Evaluation Of The RIS-Induced Pattern}\label{RIS-ind}
	{After some straightforward calculations, $\bSigma$ can be expressed as}
	$$\bSigma = \bGamma\odot \bS_1\odot\bS_2=\bv_1 \bv_2^T ,$$
	with
	\begin{eqnarray}
		\begin{aligned}
			\bv_1 = & [ e^{-j2 \pi \frac{(N-1)}{2} \Delta \kappa_u},\ldots, e^{j2 \pi\frac{(N-1)}{2} \Delta \kappa_u}]^T ,
		\end{aligned}
	\end{eqnarray}
	\begin{eqnarray}
		\begin{aligned}
			\bv_2= & [e^{-j2\pi\frac{(M-1)}{2} \Delta \kappa_v},\ldots,  e^{j2\pi\frac{(M-1)}{2} \Delta \kappa_v}]^T .
		\end{aligned}
	\end{eqnarray}
	
	As a consequence
	$$|{\bf 1}_N^T\: \bSigma \:{\bf 1}_M|^4=|{\bf 1}_N^T\bv_1|^4|{\bf 1}_M^T\bv_2|^4,$$
	with
	$$|{\bf 1}_N^T\bv_1|^4=\left|\!\frac{\sin(\Delta \kappa_u\pi N)}{\sin(\Delta \kappa_u\pi)}\!\right|^4$$
	and
	$$|{\bf 1}_M^T\bv_1|^4=\left|\!\frac{\sin(\Delta \kappa_v\pi M)}{\sin(\Delta \kappa_v\pi)}\!\right|^4  . $$

	\bibliographystyle{IEEEbib}
	\bibliography{IEEEabrv,references}
	
\end{document}